\documentclass[a4paper,twocolumn,pra,showpacs,floats,floatfix]{revtex4}
\usepackage{amsfonts}
\usepackage{amssymb}
\usepackage{amsmath}
\usepackage{graphicx}

\begin{document}

\title{Three-dimensional solitons in coupled atomic-molecular Bose-Einstein condensates}
\author{T. G. Vaughan}
\affiliation{ARC Centre of Excellence for Quantum-Atom Optics, School of Physical
Sciences, \\
The University of Queensland, Brisbane, Queensland 4072, Australia}
\author{K. V. Kheruntsyan}
\affiliation{ARC Centre of Excellence for Quantum-Atom Optics, School of Physical
Sciences, \\
The University of Queensland, Brisbane, Queensland 4072, Australia}
\author{P. D. Drummond}
\affiliation{ARC Centre of Excellence for Quantum-Atom Optics, School of Physical
Sciences, \\
The University of Queensland, Brisbane, Queensland 4072, Australia}
\date{\today{}}

\begin{abstract}
We present a theoretical analysis of three-dimensional ($3D$) matter-wave
solitons and their stability properties in coupled atomic and molecular
Bose-Einstein condensates (BEC). The soliton solutions to the mean-field
equations are obtained in an approximate analytical form by means of a
variational approach. We investigate soliton stability within the parameter
space described by the atom-molecule conversion coupling, atom-atom $s$-wave
scattering, and the bare formation energy of the molecular species. In terms
of ordinary optics, this is analogous to the process of sub/second-harmonic
generation in a quadratic non-linear medium modified by a cubic
nonlinearity, together with a phase mismatch term between the fields. While
the possibility of formation of multidimensional spatio-temporal solitons in
pure quadratic media has been theoretically demonstrated previously, here we
extend this prediction to matter-wave interactions in BEC systems where
higher-order non-linear processes due to interparticle collisions are
unavoidable and may not be neglected. The stability of the solitons
predicted for repulsive atom-atom interactions is investigated by direct
numerical simulations of the equations of motion in a full $3D$ lattice.
Our analysis also leads to a possible technique for demonstrating the ground
state of the Schr\"{o}dinger-Newton and related equations that describe
Bose-Einstein condensates with non-local inter-particle forces.
\end{abstract}

\pacs{03.75.-b, 03.75.Lm, 42.65.Tg}
\maketitle

\section{Introduction}

Recent developments in Bose-Einstein condensation of alkali gases include
the possibility of coherent molecule formation \cite{DKH98,KD98}, or
superchemistry \cite{HWDK2000} via Bose-enhanced chemical reaction at
ultra-cold temperatures. The relevant parametric quantum field theories \cite
{DKH98,KD98,Timmermans,Javanainen-Mackie-1999,KD-nondegenerate} and their
classical non-linear optical analogs \cite
{Kanashov-Rubenchik,Hillery-Mlodinov,Raymer-Drummond-etal,KD-nondegenerate}
are the subject of much current attention, due to the possibility of stable,
bright, higher-dimensional solitons (also referred to as solitary waves)
\cite{Kanashov-Rubenchik,Malomed,DKH98,DKH-JOptB-1999}. Parametric solitons
in nonlinear optics with quadratic nonlinearity have now been observed
experimentally in two transverse dimensions, both as spatial and temporal
solitons \cite{Wise-exp}.

The dynamical equations for parametric solitons are an example of
a classically non-integrable field theory which generally needs to
be treated numerically. The formation of three-dimensional ($3D$)
localized solitons is a subject of much intrinsic interest in
mathematical and non-linear physics, and it is intriguing that no
integrable models supporting them appear to exist. Despite the
absence of integrability, the quadratically coupled equations of
parametric nonlinear optics and coherently coupled
atomic-molecular Bose-Einstein condensate (BEC) systems appear to
be the simplest physically relevant Hamiltonian models having $3D$
localized solitons \cite{Ruostekoski}.

They also provide an experimental route towards demonstration of the closely
related ground state of the Schr\"{o}dinger-Newton (SN)\ equation \cite{SN}
introduced to describe gravitationally bound Bose gases, and later revived
by Penrose and others \cite{SN-Penrose,SN-2} as a possible model of the
collapse of the quantum-mechanical wave-function. We also show that there
are parallels with more general mean-field models of Bose gases having a
combination of short-distance repulsion and finite-range Yukawa-like
attractive interactions. These more general models may also have
astrophysical or quantum-mechanical significance.

The surprisingly close parallels to nonlinear optics, in the related fields
of quantum many-body theory and atom optics, have now resulted in the
emergence of a new field of research -- nonlinear atom optics with
parametric nonlinearity. The first step towards seeing molecular
condensation was recently undertaken in transient experiments with a
Bose-Einstein condensate of $^{85}$Rb atoms \cite{Donley}, in which
interference measurements were indicative of coherent molecule formation.
More recent experiments with $^{133}$Cs, $^{87}$Rb, and $^{23}$Na \cite
{Bosonic-Cs-Rb-Na}, as well as with degenerate Fermi gases of $^{40}$K and $
^{6}$Li atoms \cite{Fermionic}, have produced even larger fraction of
ultra-cold molecules, as well as Bose-Einstein condensates of bosonic
molecular dimers composed of fermionic atoms. All these experiments have
employed magnetic Feshbach resonances, which appear to be more successful at
present than the alternative Raman photo-association scheme \cite
{Heinzen-exp}.

In addition to this remarkable experimental progress, the original effective
quantum field theory \cite
{DKH98,KD98,Timmermans,Javanainen-Mackie-1999,KD-nondegenerate,HWDK2000} for
coupled atomic-molecular BECs has also been developed; it now incorporates
renormalization, the treatment of intrinsic pair correlations, quantum
fluctuations, and thermal effects (see, e.g., \cite
{Verhaar,Yurovski-Julienne,Holland-2001,Goral-Rzazewski-2001,Hope-Olsen,Kokkelmans-Holland-2002,Mackie-Suominen-Javanainen-2002,Koehler-Burnett-2003,Duine-Stoof-JOptB-PRL}, as well as a recent review paper \cite{Duine-Stoof-review} for further
references).

In the near-classical limit of large numbers of atoms or photons, the
relevant equations for parametric solitons are those of mean-field theory,
which is a modified version of the Gross-Pitaevskii (GP) equation (in the
atomic BEC case) or the nonlinear Schr\"{o}dinger equation (in the photonic
case). These are in fact identical equations, except expressed in the
different languages of condensed matter physics or photonics. The
modification consists of the addition of a parametric nonlinear term
analogous to a quadratic nonlinearity in nonlinear optics. This couples the
atomic and molecular fields together by means of a coherent inter-conversion
process. The parametric coupling acts as an energy-lowering `glue', which
can permit stable mutually-trapped BEC solitons to form in $3D$. In the
absence of gravity, this would imply the possibility of localized matter
waves in free space without external trapping potentials. Unlike the usual
GP equations, the existence of attractive forces in three dimensions does
not result in a catastrophic collapse, provided the $s$-wave scattering
length is non-negative.

In nonlinear optics, various aspects of the competition between quadratic
and cubic nonlinearities on soliton formation have been studied in Refs.
\cite{parametric-quartic-interplay}, either in lower dimensions ($1D$ or $2D$
) or in cases of special relations between the system parameters.
Matter-wave solitons in $3D$ coupled atomic-molecular BECs have been studied
in Refs. \cite{DKH98,DKH-JOptB-1999,KD-nondegenerate} and \cite
{Kivshar-AMBEC-solitons}, but only for specific values of the $s$-wave
couplings.

In the present paper, we extend these results to the case of arbitrary
interaction strengths for the atom-molecule coupling and for the repulsive
atom-atom $s$-wave scattering, as well as for arbitrary energy mismatch
between the atoms and molecules. We analyze the superchemistry equations in
the mean-field limit, to obtain the precise conditions under which $3D$
atomic-molecular BEC solitons can form. Approximate soliton solutions are
found analytically, by means of a variational approach with a Gaussian and
an exponential ansatz. We then numerically study the dynamical stability of
the resulting solitons on a $3D$ lattice, together with comparing the
results with exact numerically solutions.

We find that there are large regions of stability in parameter space,
depending on the energy difference between the atomic and molecular
condensates, the numbers of atoms involved, and the coupling strengths. For
simplicity, the analysis only includes repulsive $s$-wave scattering between
the atoms, and assumes no other $s$-wave interactions. While more general $s$
-wave interactions are simple enough to include, we have focused on a
relatively simple case here in the interest of keeping the parameter-space
manageable.

\section{The model}

We start with an effective field theory model for a coherently coupled
atomic-molecular system \cite{DKH98}. The model and the obtained results can
easily be adopted to describe certain cases of non-linear optical
interactions of second- and sub-harmonic waves in a nonlinear crystal \cite
{KD98,KD-nondegenerate}. In the atom-molecular case, the model refers to a
type of superchemistry \cite{HWDK2000}, in which an atomic condensate is
able to coherently and reversibly inter-convert with a condensate of
diatomic molecules.

There are several possible experimental routes for providing this type of
coherent coupling, including a Feshbach resonance (employing a tuned
external DC magnetic field), Raman photo-association (involving two external
lasers with a well-defined frequency difference), and direct single-photon
photo-association (this would require an external microwave or infra-red
field) \cite
{DKH98,KD98,HWDK2000,Timmermans,Javanainen-Mackie-1999,Verhaar,Yurovski-Julienne,Holland-2001,Goral-Rzazewski-2001,Hope-Olsen,Kokkelmans-Holland-2002,Mackie-Suominen-Javanainen-2002,Koehler-Burnett-2003,Duine-Stoof-JOptB-PRL,Duine-Stoof-review}. The first two cases have been experimentally demonstrated \cite
{Heinzen-exp,Donley}, although not yet the last. In practical terms, the
main limitation of the model and corresponding experiments is the need to
minimize incoherent processes like inelastic collisions and other loss
processes.

\subsection{Hamiltonian}

In the model, we suppose that each condensate has the usual kinetic energy
term, atom-atom $s$-wave scattering interactions, and a number-conserving
coherent coupling of the form $\hat{\Phi}^{\dagger }\hat{\Phi}^{\dagger }
\hat{\Psi}$, where $\hat{\Phi}$ represents the atomic field, and $\hat{\Psi}$
is the field operator for the molecular dimers. In $D$ ($D=1,2,3$) spatial
dimensions, this leads to a model Hamiltonian of the following form:
\begin{align}
\hat{H}& =\int d^{D}\mathbf{x}\left[ \frac{\hbar ^{2}}{2m_{1}
}|\nabla \hat{\Phi}(\mathbf{x})|^{2}+\frac{\hbar ^{2}}{2m_{2}}|\nabla \hat{
\Psi}(\mathbf{x})|^{2}\right.   \notag \\
& \left. +V_{\Psi }\hat{\Psi}^{\dag }(\mathbf{x})\hat{\Psi}(\mathbf{x}
) +\frac{\hbar \kappa _{11}}{2}\hat{\Phi}^{\dagger }(\mathbf{x})\hat{
\Phi}^{\dagger }(\mathbf{x})\hat{\Phi}(\mathbf{x})\hat{\Phi}(\mathbf{x})
\right.   \notag \\
& \left. -\,\frac{\hbar \chi }{2}\left( \hat{\Phi}^{\dagger }(\mathbf{x})
\hat{\Phi}^{\dagger }(\mathbf{x})\hat{\Psi}(\mathbf{x})+\hat{\Psi}^{\dagger
}(\mathbf{x})\hat{\Phi}(\mathbf{x})\hat{\Phi}(\mathbf{x})\right)
\right] .  \label{H(Phi,Psi)}
\end{align}
Here, $m_{1}$ and $m_{2}$ are the masses of the atoms and
molecules, respectively, $V_{\Psi }$ is the internal molecular
energy relative to free atoms, and the coupling $\chi $ (which we
assume is positive) describes coherent conversion of pairs of
atoms into diatomic molecules and vice versa. The atomic
self-interaction strength, $\kappa_{11}$, is proportional to the
$s$-wave scattering length. For example, in $3D$,
$\kappa_{11}=4\pi \hbar a_{11}/m_{1}$, where $a_{11}$ is the
atom-atom scattering length which is assumed positive, as is
usually needed to form a stable BEC in the first place.

To allow comparisons with the Schr\"{o}dinger-Newton (SN) \cite
{SN,SN-Penrose,SN-2} equation, we will also consider a related
model in which the interaction term $-\,\hbar \chi \lbrack \hat{\Phi}
^{\dagger }\hat{\Phi}^{\dagger }\hat{\Psi}+\hat{\Psi}^{\dagger }\hat{\Phi}
\hat{\Phi}]/2$ is replaced by $-\,\hbar \chi \hat{\Phi}^{\dagger }\hat{\Phi}[
\hat{\Psi}+\hat{\Psi}^{\dagger }]/2$. This models a Bose-Einstein condensate
$\hat{\Phi}(\mathbf{x})$ with short-range interactions, together with a
long-range attractive force caused by the exchange of a meson-like particle $
\hat{\Psi}(\mathbf{x})$. In the mean-field theory limit, we call this model
the Gross-Pitaevskii-Yukawa or GPY model, which is more general than the
Schr\"{o}dinger-Newton model.

In astrophysical situations, the GPY mean-field theory reduces to
the SN model in the combined limit of zero $s$-wave scattering,
and an infinitely long-range gravitational interaction with $
m_{2}\rightarrow 0$, leading to an inverse-square law \cite{SN}.
The presence of a short-range interaction $s$-wave scattering term
makes the GPY model more realistic than the usual SN model. The SN
model is used to describe a degenerate Bose gas with gravitational
self-interaction, and has also been suggested as a possible
mechanism for wave-packet collapse in quantum mechanics
\cite{SN-Penrose,SN-2}. In the one-dimensional case, the GPY
theory is similar to the nonlinear interactions in an optical
fiber caused by couplings of photons to phonons \cite{CD}. At the
same time, the GPY model allows one to investigate attractive
interactions with more general behavior than a simple
inverse-square law.

At the quantum field level, either model Hamiltonian implicitly
involves a delta-function effective interaction between the atoms,
and so requires the use of a momentum cutoff $k_{max}\ll 1/a_{11}$
for selfconsistency. More rigorous regularization consists of
renormalization of the theory \cite {Kokkelmans-Holland-2002} as
$k_{max}\rightarrow \infty $. In the present paper, however, we
employ a mean-field approximation, and restrict ourselves to the
study of solitons that have spatial widths much larger than
$k_{max}^{-1}$ so that the cutoff dependencies are negligible,
while the relevant parameters are the \emph{observed} couplings.
The mean-field theory is also a high-density approximation, since
quantum fluctuations and correlations are expected to cause quite
different ground state to appear at low density
\cite{DKH98,KD98,Holland-2001,Hope-Olsen}.

A more complete model Hamiltonian should also incorporate
atom-molecule and molecule-molecule $s$-wave scattering
interactions. However, these greatly complicate the analysis,
without adding much qualitatively new physics to the $3D$ soliton
properties studied here. In addition, the respective scattering
lengths are not known yet in most cases. For this reason, we
assume that the atom-atom scattering is the dominant $s$-wave
interaction and simply omit all other $s$-wave scattering
processes, in the interest of simplicity.

\subsection{Mean-field equations}

The corresponding equations of motion for the mean fields in the
atom-molecular model, following from the Hamiltonian (\ref{H(Phi,Psi)}) and
valid at high densities, are
\begin{align}
i\frac{\partial\phi(\mathbf{x},t)}{\partial t} & =-\frac{\hbar}{2m_{1} }
\nabla_{\mathbf{x}}^{2}\phi-\chi\phi^{\ast}\psi+\kappa_{11}|\phi|^{2} \phi,
\notag \\
i\frac{\partial\psi(\mathbf{x},t)}{\partial t} & =-\frac{\hbar}{2m_{2} }
\nabla_{\mathbf{x}}^{2}\psi+\Delta\omega\psi-\frac{1}{2}\chi\phi^{2},
\end{align}
where $\hbar\Delta\omega=V_{\psi}$ is the energy mismatch on converting
atoms to molecules, and $m_{2}=2m_{1}$.

In this model, the total number of particles $N$ (i.e. the total number of
atomic particles, including pairs of atoms inside the diatomic molecules)
here is conserved:
\begin{equation}
N=N_{1}+2N_{2}=\int d^{D}\mathbf{x}\left[ {\,}|\phi(\mathbf{x},t)|^{2}
+2|\psi(\mathbf{x},t)|^{2}\right].  \label{eqn:N_total}
\end{equation}

For completeness, we include the mean-field equations for the related GPY
equations. These have the structure:
\begin{align}
i\frac{\partial\phi(\mathbf{x},t)}{\partial t} & =-\frac{\hbar}{2m_{1}}
\nabla_{\mathbf{x}}^{2}\phi-\frac{\chi}{2}\phi(\psi+\psi^{\ast})+\kappa
_{11}|\phi|^{2}\phi,  \notag \\
i\frac{\partial\psi(\mathbf{x},t)}{\partial t} & =-\frac{\hbar}{2m_{2}}
\nabla_{\mathbf{x}}^{2}\psi+\Delta\omega\psi-\frac{1}{2}\chi|\phi|^{2}.\,\,
\end{align}

In the limit of $m_{2}\rightarrow0$, $\chi\rightarrow\infty$, so
that $ \Delta\omega m_{2}\rightarrow0$ and $\chi^{2}m_{2}=4\pi
Gm_{1}^{2}$, and assuming that $\psi$ is real, we introduce a
gravitational field potential $V_{g}=-\chi\psi$. In this
long-range force limit, one can apply an adiabatic approximation
to the second equation, which leads to the Poisson equation:
\begin{align}
i\hbar\frac{\partial\phi(\mathbf{x},t)}{\partial t} & =-\frac{\hbar^{2} }{
2m_{1}}\nabla_{\mathbf{x}}^{2}\phi+V_{g}\phi+\hbar\kappa_{11}|\phi|^{2} \phi,
\notag \\
\nabla_{\mathbf{x}}^{2}V_{g} & =4\pi Gm_{1}^{2}|\phi|^{2}.\,\,
\label{timedependentSN}
\end{align}

This can be recognized as the mean-field equation for a BEC having
an additional self-gravitational force with the gravitational
potential energy $V_{g}$ and gravitational constant $G$, as well
as the usual GP short-range interaction. Here, the conserved
particle number is given by $N=\int d^{3} \mathbf{x}{\,}|\phi
(\mathbf{x},t)|^{2}$. In the additional limit of $\kappa
_{11}\rightarrow 0$, the equations correspond to the
time-dependent version of the SN equation
\cite{SN,SN-Penrose,SN-2} in which there is no short-range
self-interaction.

\subsection{Dimensionless variables}

\subsubsection{Atom-molecular system}

In general, atom-molecular solitons may exist with periodic phase and
frequency $\omega $, so that $i\partial \phi/\partial t =\omega \phi$
and $i\partial \psi/\partial t =2\omega \psi$.

We introduce a characteristic time scale $t_{0}$ and a characteristic
length scale $d_{0}=\sqrt{\hbar t_{0}/2m_{1}}$. We also transform to
dimensionless time and position variables,
\begin{align}
\tau & =t/t_{0},  \notag \\
\xi_{i} & =x_{i}/d_{0}, \label{eqn:tau-xi}
\end{align}
and dimensionless fields,
\begin{align}
u & =\chi t_{0}\phi e^{i\omega t},  \notag \\
v & =\chi t_{0}\psi e^{2i\omega t}.
\end{align}

This gives the corresponding equations of motion in dimensionless form, with
no reduction in parameter space:
\begin{align}
i\frac{\partial u}{\partial\tau} & =-\nabla_{\mathbf{\xi}}^{2}u+\gamma
_{u}u-u_{{}}^{\ast}v+\alpha_{11}|u|^{2}u,  \notag \\
i\frac{\partial v}{\partial\tau} & =-\frac{1}{2}\nabla_{\mathbf{\xi}} ^{2}v+
\frac{\gamma}{2}v-\frac{1}{2}u^{2}\,\,.  \label{eqn:nodim-eom}
\end{align}

Here we have introduced new dimensionless parameters according to
\begin{align}
\alpha_{11} & =\frac{\kappa_{11}}{\chi^{2}t_{0}},  \notag \\
\gamma_{u} & =-\omega t_{0},  \notag \\
\gamma & =(2\Delta\omega-4\omega)t_{0}.
\end{align}

To make the scaling definite, we can set $\gamma_{u}=1$ with no loss in
generality, provided $\omega<0$. This corresponds to a localized bound state
with negative energy $E=\hbar\omega$; we do not investigate the unbound
solutions here. The choice $\gamma_{u}=1$ also gives a simple relationship
between the dimensionless parameter $\gamma$ and the detuning
$\Delta\omega$,
\begin{equation}
\gamma=4+2\Delta\omega t_{0},  \label{gamma-Deltaomega}
\end{equation}
corresponding to a shifted energy mismatch.

For $\alpha_{11}=0$ (and with an additional scaling of $u\rightarrow u/\sqrt{
2}$), Eqs. (\ref{eqn:nodim-eom}) are equivalent to Eqs. (1) and (2) of Ref.
\cite{Malomed}, with the value of the coefficient $\delta=1$. This
corresponds to optical parametric interaction in a quadratically nonlinear
medium, in which the dispersion coefficients for the fundamental and
second-harmonic fields are equal to each other.

We seek stationary solutions ($\partial u/\partial\tau=\partial
v/\partial\tau=0$) to the equations of motion (\ref{eqn:nodim-eom}), i.e.
those that have
\begin{align}
\nabla_{\mathbf{\xi}}^{2}u & =u-u_{{}}^{\ast}v+\alpha_{11}|u|^{2} u,  \notag
\\
\nabla_{\mathbf{\xi}}^{2}v & =\gamma v-u^{2}.  \label{udotvdot0}
\end{align}
These correspond to extrema of the following dimensionless atom-molecular
Hamiltonian
\begin{align}
H^{(u,v)} & =\int d^{D}\mathbf{\xi}\left[ |\nabla u|^{2}+\frac{1}{2}|\nabla
v|^{2}+|u|^{2}+\frac{\gamma}{2}|v|^{2}\right.  \notag \\
& \left. -\,\frac{1}{2}\left( (u^{\ast})^{2}v+c.c.\right) +\frac {\alpha_{11}
}{2}|u|^{4}\right],  \label{Huv}
\end{align}
where the expression for the original Hamiltonian energy in terms of the
dimensionless variables is
\begin{equation}
H=\left( \frac{\hbar}{2m_{1}t_{0}}\right) ^{D/2}\frac{\hbar }{{
\chi^{2}}}H^{(u,v)}\,.
\end{equation}

A conserved quantity for the ($u,v$) system, which is proportional to
the total number of particles $N$, is
\begin{equation}
\mathcal{N}^{\prime}=\int d^{D}{\mathbf{\xi}}\left[ |u(\mathbf{\xi}
,t)|^{2}+2|v(\mathbf{\xi},t)|^{2}\right] .  \label{N-prime}
\end{equation}

\subsubsection{Schr\"{o}dinger-Newton system}

Similarly, stationary solutions to the time-dependent SN equation
(\ref{timedependentSN}) may also exist with frequency $\omega$,
so that $i\partial \phi/\partial t = \omega\phi$,
with $V_{g}$ following adiabatically. This translates
Eq.~(\ref{timedependentSN}) directly to the time-independent SN
equation appearing in
Refs.~\cite{SN,SN-2}, for $\phi_{s}=\phi\exp(i\omega t)$:
\begin{align}
\frac{\hbar^{2}}{2m_{1}}\nabla_{\mathbf{x}}^{2}\phi_{s} &
=-E\phi_{s}+V_{g}\phi_{s},  \notag \\
\nabla_{\mathbf{x}}^{2}V_{g} & =4\pi Gm_{1}^{2}|\phi_{s}|^{2},
\end{align}
where $E=\hbar\omega$.

By introducing characteristic time and length scales, as in
Eqs.~(\ref{eqn:tau-xi}), and transforming to dimensionless fields
\begin{align}
u &= t_{0}\sqrt{2\pi Gm_{1}}\phi_{s}, \notag \\
v &= -\frac{t_{0}}{\hbar}V_{g},
\end{align}
we obtain the dimensionless time-independent SN equations:
\begin{align}
\nabla_{\mathbf{\xi}}^2 u &= u - vu, \notag \\
\nabla_{\mathbf{\xi}}^2 v &= -u^2.
\label{stationarySN}
\end{align}
The normalization is now
given by $\int d^{D}\mathbf{\xi}{\,}u^{2}=|\mathcal{E}|^{-1/2}$. Here
$\mathcal{E} =E/E_{0}$ is the dimensionless energy ($\mathcal{E}<0$), and the
energy scale $E_{0}$ is defined via $E_{0}=32\pi^{2}m_{1}^{5}N^{2}G^{2}/
\hbar^{2}$.

Note that Eqs.~(\ref{stationarySN}) are identical to Eqs.~(\ref{udotvdot0})
for real stationary solutions to the atom-molecular system when
$\gamma=\alpha_{11}=0$.

\section{Variational analysis}

\subsection{Gaussian variational ansatz (GVA)}

To analyze the localized soliton solutions to Eqs. (\ref{eqn:nodim-eom}),
where $\gamma _{u}=1$, we introduce an approximate Gaussian variational
ansatz (GVA). This permits an analytic treatment of the problem of
minimizing the Hamiltonian energy. The stability of the GVA solutions will
be checked numerically by dynamical evolution of the equations of motion
where the GVA serves as the initial condition.

The GVA solutions are introduced according to
\begin{align}
u(\xi,\tau=0) & =Ae^{-a\xi^{2}},  \notag \\
v(\xi,\tau=0) & =Be^{-b\xi^{2}},
\end{align}
where $\xi=|\mathbf{\xi}|$. Here, the parameters $a$ and $b$ must both be
real and positive for localized solitons, and we assume that both the
amplitudes $A$ and $B$ are also real and positive. This choice in fact
already takes care of the optimum relative phase between the atomic and
molecular fields, where we can without loss of generality take $B$ to be
real, while the optimum relative phase will dictate the phase of $A$. This
immediately leads us to the conclusion that $A^{2}$ must be real and
positive too, in order that the atom-molecule interaction term remains
negative (for positive $\chi$ as assumed here) and permits a minimum in the
Hamiltonian energy. The sign of $A$ is in fact irrelevant, since the
corresponding equations of motion (\ref{eqn:nodim-eom}) are invariant under
the sign change of $A$.

Substituting the GVA into Eqs. (\ref{N-prime}) and (\ref{Huv}) for
$\mathcal{N}^{\prime}$ and the Hamiltonian energy $H^{(u,v)}$, and
taking the integrals we obtain, in $D=1,2$ or $3$ space
dimensions:
\begin{equation}
\mathcal{N}^{\prime}=\left( \frac{\pi}{2}\right) ^{D/2}\left[ \frac{A^{2} }{
a^{D/2}}+\frac{2B^{2}}{b^{D/2}}\right] ,  \label{N-prime(A,B,a,b)}
\end{equation}
\begin{align}
H^{(u,v)} & =\frac{D}{2}\left( \frac{\pi}{2}\right) ^{D/2} \left[
\frac{2A^{2}}{a^{D/2-1}}+\frac{B^{2}}{b^{D/2-1}}+\frac {2A^{2}
}{Da^{D/2}}\right.  \notag \\
& \left. +\frac{\gamma B^{2}}{Db^{D/2}}
-\frac{2^{1+D/2}A^{2}B}{D(2a+b)^{D/2}}+\frac{\alpha_{11}A^{4} }{
D(2a)^{D/2}}\right] .  \label{H-prime(A,B,a,b)}
\end{align}

Minimizing $H^{(u,v)}$ with respect to $a,b,A,$ and $B$ (for
given $\alpha_{11}$ and $\gamma$) gives the following solution:
\begin{equation}
a=\frac{b}{2}\left[ \frac{2(Db+\gamma)}{(D-2)b+\gamma}-1\right] ,  \label{a}
\end{equation}
\begin{align}
& \alpha_{11}(2a+b)^{1+D}(Db+\gamma)\left[ (4-D)a-1\right]  \notag \\
& -\left. 2^{1+D}(2ab)^{D/2}\left[ 4a^{2}-(D-2)ab-b\right] =0\right. ,
\label{b}
\end{align}
\begin{align}
B & =\frac{(2a+b)^{D/2}(Da+1)}{(2a)^{D/2}}  \notag \\
& \times\left[ 1-\frac{\alpha_{11}(2a+b)^{D}(Db+\gamma)}{2^{D}(2ab)^{D/2} }
\right] ^{-1},  \label{B}
\end{align}
\begin{equation}
A^{2}=\frac{(2a+b)^{D/2}(Db+\gamma)}{(2b)^{D/2}}B,  \label{A}
\end{equation}
In Eq. (\ref{b}), the parameter $a$ is to be substituted using
Eq.~(\ref{a}), so that Eq. (\ref{b}) (to be solved first) reads as
a polynomial equation with respect to $b$, for given values of
$\alpha_{11}$ and $\gamma$. Alternatively, $b$ can be regarded as
a free parameter and Eq. (\ref{b}) be viewed and easily solved
with respect to $\alpha_{11}$, for given $\gamma$ and $b$. In
doing so, only positive $b$-values have to be considered for
physically meaningful (localized) soliton solutions.

For certain values of $\alpha_{11}$ and $\gamma$ this system has a unique
solution (see Sec. IV A), giving the soliton parameters $A,B,a$ and $b$. The
soliton parameters give in turn the resulting value of the conserved
quantity $\mathcal{N}^{\prime}$, Eq. (\ref{N-prime(A,B,a,b)}), and the
Hamiltonian energy (\ref{H-prime(A,B,a,b)}).

\subsection{Exponential variational ansatz (EVA)}

It can be shown that any localized stationary solution to
the equations of motion (\ref{eqn:nodim-eom}) (with $\gamma_{u}=1$) must
possess tails decaying according to:
\begin{align}
u(\xi\gg1,\tau) & \propto e^{-\xi}/\xi,  \notag \\
v(\xi\gg1,\tau) & \propto e^{-\sqrt{\gamma}\xi}/\xi,  \label{exp-tails}
\end{align}
where $\xi=|\mathbf{\xi}|$. This result can be obtained by neglecting all non-linear
terms in Eqs. (\ref{eqn:nodim-eom}) at large $\xi$, and solving the
resulting decoupled equations for the stationary states.

Due to the singularity at origin, direct employment of this
variational trial function would be problematic. However, since it
indicates that the soliton tails should decay slower than those of
the Gaussian trial functions, we are motivated to also consider an
alternative \emph{exponential} variational ansatz (EVA):
\begin{align}
u(\xi ,\tau & =0)=Pe^{-p\sqrt{\xi ^{2}+\epsilon }},  \notag \\
v(\xi ,\tau & =0)=Qe^{-q\sqrt{\xi ^{2}+\epsilon }}.
\end{align}
As in the case of the GVA, here too we assume that $p$, $q$, $P$
and $Q$ are all real and positive. For analytic simplicity, we let
$\epsilon$ be an infinitely small length scale, which is formally
included to ensure that $u$ and $v$ are differentiable at $\xi
=0$. We then proceed and evaluate the integrals in
$\mathcal{N}^{\prime }$ and $H^{(u,v)}$ to find that, as $\epsilon
\rightarrow 0$:
\begin{equation}
\mathcal{N}^{\prime }=K_{D}\left[ \frac{P^{2}}{p^{D}}+\frac{2Q^{2}}{q^{D}}
\right],
\end{equation}
\begin{align}
H^{(u,v)}& =K_{D}\left[ \frac{P^{2}}{p^{D-2}}+\frac{Q^{2}}{2q^{D-2}}+\frac{
P^{2}}{p^{D}}+\gamma \frac{Q^{2}}{2q^{D}}\right.   \notag \\
& \left. -\frac{2^{D}P^{2}Q}{(2p+q)^{D}}+\alpha _{11}\frac{P^{4}}{
2^{D+1}p^{D}}\right] .
\end{align}
Here, $K_{D}=1,\pi /2,\pi $ for $D=1,2,3$,
respectively.

Variational stationary points are then given by the solution to the
following set of algebraic equations:
\begin{equation}
p=\frac{q}{2}\left[ \frac{2D(q^{2}+\gamma)}{(D-2)q^{2}+D\gamma}-1\right],
\label{eqn:p}
\end{equation}
\begin{align}
& \!\!\!\!\!\!\!\!2^{3D+1}(pq)^{D}\left\{ \left[ (D-2)p^{2}+D\right]
(2p+q)-2Dp(p^{2}+1)\right\}  \notag \\
& -\alpha_{11}(2p+q)^{2D+1}\left( q^{2}+\gamma\right) \left[ D-(4-D)p^{2}
\right] =0,  \label{eqn:q}
\end{align}
\begin{align}
Q & =\frac{((D-2)p^{2}+D)(2p+q)^{D+1}}{D(2p)^{D+1}}  \notag \\
& \times\left[ 1-\alpha_{11}\frac{(q^{2}+\gamma)(2p+q)^{2D+1}} {
2^{3D+2}p^{D+1}q^{D}}\right] ^{-1},  \label{eqn:Q}
\end{align}
\begin{equation}
P^{2}=\frac{(q^{2}+\gamma)(2p+q)^{D}}{(2q)^{D}}Q.  \label{eqn:P}
\end{equation}

To compare the EVA solutions with those of the GVA, it is necessary to
ensure that both have identical $\mathcal{N}^{\prime }$ for a given pair of
the parameters $\alpha _{11}$ and $\gamma $. As the solutions given above
result from (unconstrained) variational minimization with respect to all
parameters, this requirement will not in general be met. We thus
perform a constrained minimization with respect to $p$, $q$ and $Q$, leaving
$P$ to be fixed by $\mathcal{N}^{\prime }$ of the associated GVA solution
\cite{constrained-variables}. The constrained EVA solution for $D=3$ is then
given by:
\begin{equation}
\frac{3}{8}\alpha _{11}pF-\frac{48pqQ}{(2p+q)^{4}}+4=0,
\label{eqn:constrained_EVA1}
\end{equation}
\begin{align}
& F\left[ \frac{3}{2}\alpha _{11}p^{3}Q+48\frac{p^{3}q^{4}}{(2p+q)^{4}}
\right]   \notag \\
& +12Q\left[ 1+p^{2}-\frac{8p^{3}Q}{(2p+q)^{3}}\right] -Q(3\gamma
+q^{2})=0,  \label{eqn:constrained_EVA2}
\end{align}
\begin{align}
& F\left[ \alpha _{11}\frac{p^{3}}{q^{3}}+\frac{16p^{3}}{(2p+q)^{3}}\right]
\notag \\
& +\frac{8Q}{q^{3}}\left[
1+p^{2}-\frac{8p^{3}Q}{(2p+q)^{3}}\right]    -\frac{2Q}{b}\left[
1+\frac{\gamma }{q^{2}}\right] =0, \label{eqn:constrained_EVA3}
\end{align}
where we have defined $F\equiv \mathcal{N}^{\prime }/\pi -2Q^{2}/q^{3}$.

We note that a similar exponential variational solution for the `atomic' $u$
-field (though not for the `molecular' $v$-field) has been used previously
for the Schr\"{o}dinger-Newton equation \cite{SN-2}.

\section{$3D$ soliton properties}

\subsection{Existence and properties of GVA solutions}

\begin{figure}[ptb]
\includegraphics[height=6cm]{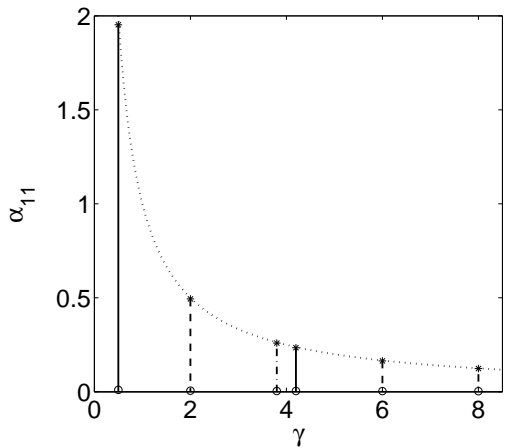}
\caption{GVA solutions existence domain in the ($\protect\gamma,\protect
\alpha_{11}$) plane, for $\protect\gamma>0$. The dotted line gives the upper
bound on $\protect\alpha_{11}$ corresponding to the boundary $\protect\alpha
_{11}<1/\protect\gamma$. The set of vertical lines in the region $\protect
\gamma<4$ ($\protect\gamma=0.5$ -- full line, $\protect\gamma=2$ -- dashed
line, and $\protect\gamma=3.8$ -- dash-dotted line), and in the region $
\protect\gamma>4$ ($\protect\gamma=4.2$ -- full line, $\protect\gamma=6$ --
dashed line, and $\protect\gamma=8$ -- dash-dotted line) are to serve for
mapping purposes as discussed in the text and explained in the captions to
subsequent figures.}
\label{fig:GVA_domain}
\end{figure}

In order to analyze the properties of the GVA solutions,
Eqs.~(\ref{a})-(\ref{A}), for a given pair of $\gamma$ and
$\alpha_{11}$, we first note that our analysis is restricted to
the case of $\alpha_{11}\geq0$, i.e. repulsive atom-atom
interactions including a non-interacting limit of $\alpha_{11}=0$.
In addition, we restrict ourselves to the three-dimensional case
($D=3$) only.

Next, any localized physical solutions require $a$ and $b$ to be both real
and positive. The existence of the minimum for the Hamiltonian energy,
Eq.~(\ref{H-prime(A,B,a,b)}), requires that the product $A^{2}B$ is real and
positive too, so that the atom-molecule conversion term gives a negative
contribution to the energy. As we mentioned earlier, this is achieved by
taking both $A$ and $B$ to be positive.

To investigate the consequences of these requirements in terms of the
soliton existence domain in the parameter space $(\alpha_{11},\gamma)$, one
has to start from solving numerically the polynomial Eq. (\ref{b}). However,
a simpler route that allows us to obtain analytical results is to view $b$
as a free positive-valued parameter and solve Eq. (\ref{b}) for $\alpha_{11}$
in terms of $b$ and $\gamma$. In this case, the GVA solutions can be
rewritten (for $D=3$) in a simpler form:
\begin{align}
a & =\frac{b(5b+\gamma)}{2(b+\gamma)},  \label{a-3D} \\
\alpha_{11} & =\frac{2^{4+3/2}a^{3/2}b^{3/2}(4a^{2}-ab-b)}{(2a+b)^{4}
(3b+\gamma)(a-1)},  \label{alpha-3D} \\
B & =\frac{(2a+b)^{5/2}(a-1)}{2^{3/2}a^{3/2}(b-2a)},  \label{B-3D} \\
A^{2} & =\frac{(2a+b)^{3/2}(3b+\gamma)}{2^{3/2}b^{3/2}}B,  \label{A-3D}
\end{align}
where we have substituted the solution for $\alpha_{11}$ into the expression
for $B$, and the parameter $a$ in Eqs. (\ref{alpha-3D})-(\ref{A-3D}) is to
be substituted using Eq. (\ref{a-3D}).

\begin{figure}[ptb]
\includegraphics[height=7.6cm]{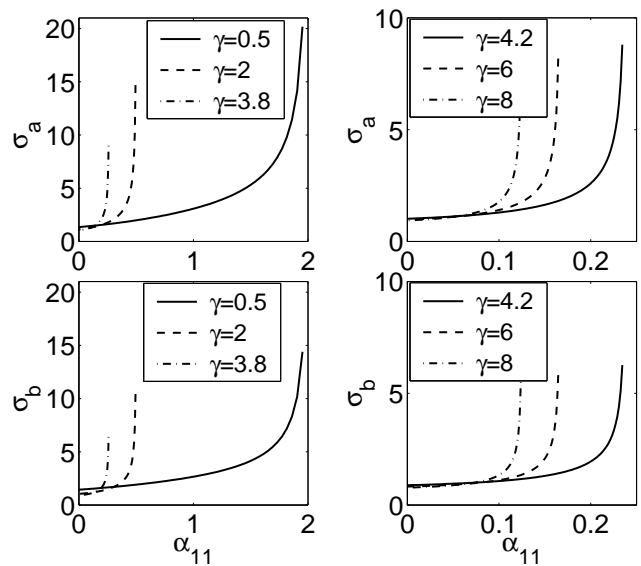}
\caption{GVA solution widths $\protect\sigma_{a}=1/\protect\sqrt{2a}$ and $
\protect\sigma_{b} =1/\protect\sqrt{2b}$ as a function of $\protect\alpha
_{11}$, for different values of $\protect\gamma$ corresponding to different
vertical lines in Fig. \protect\ref{fig:GVA_domain}.}
\label{fig:GVA_widths}
\end{figure}

In Appendix A, we analyze the above set of equations for possible solutions.
Restricting ourselves to the physically interesting subspace of $\gamma>0$,
we find that the soliton existence domain for the GVA is given by $0\leq
\alpha_{11}<1/\gamma$. The parameter space identifying this in the $
(\gamma,\alpha_{11})$-plane is shown in Fig. \ref{fig:GVA_domain}. Here, the
vertical lines at $\gamma=0.5$, $\gamma=2$, $\gamma=3.8$, $\gamma=4.2$, $
\gamma=6$, and $\gamma=8$ serve as test lines for mapping purposes discussed
in subsequent sections. The dotted line gives the upper bound on $\alpha_{11}$
corresponding to the boundary $\alpha_{11}<1/\gamma$.

\begin{figure}[ptb]
\includegraphics[height=3.9cm]{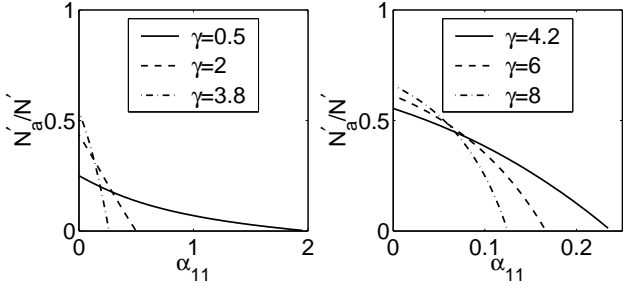}
\caption{Fraction of the number of particles in the atomic field $\mathcal{N}
_{a}^{\prime} /\mathcal{N}^{\prime}$ for GVA solutions as a function of $
\protect\alpha_{11}$, along the lines of fixed $\protect\gamma$-values as
shown in Fig. \protect\ref{fig:GVA_domain}.}
\label{fig:GVA_fraction}
\end{figure}

Figure \ref{fig:GVA_widths} represents the GVA soliton widths $\sigma _{a}=1/
\sqrt{2a}$ and $\sigma_{b}=1/\sqrt{2b}$ as a function of $\alpha_{11}$ ($
0\leq\alpha_{11}<1/\gamma$), for different values of $\gamma$. Similarly,
Fig. \ref{fig:GVA_fraction} represents the fraction of the number of
particles in the atomic field $\mathcal{N}_{a}^{\prime}/\mathcal{N}^{\prime}$
versus $\alpha_{11}$, where $\mathcal{N}_{a}^{\prime}=\int d^{3}{\xi}u^{2}$
is proportional to the total number of atoms, while $\mathcal{N}
^{\prime}=\int d^{3}{\xi(}u^{2}+2v^{2})$ is proportional to the total number
of atomic particles including pairs of atoms in the molecular component. Due
to the conserved total particle number, the fraction of molecules is found
from $\mathcal{N}_{mol}^{\prime}/\mathcal{N}^{\prime} =0.5(1-\mathcal{N}
_{a}^{\prime}/\mathcal{N}^{\prime})$.

As we can see, for large negative detuning $\Delta\omega$, corresponding to $
\gamma\ll4$ ($\gamma>0$), and for vanishing atom-atom repulsion ($\alpha
_{11}\simeq0$), the atomic fraction is relatively small and increases
monotonically with increasing $\gamma$. In all cases, the atomic fraction
decreases rapidly as $\alpha_{11}$ increases, due to the increased energy
penalty resulting from interatomic interactions. The graphs for the soliton
widths show that the atomic density profiles are in general wider than the
corresponding molecular density profiles, and that the atomic component
becomes wider and lower in the amplitude in the limit of strong inter-atomic
repulsion, $\alpha_{11}\rightarrow1/\gamma$. In this limit, it is
energetically preferable for atom pairs to populate the molecular component
so that the stable configuration of the system is a pure molecular
condensate. We note that this is a consequence of the fact that our model
neglects the molecule-molecule self-interaction completely.

\subsection{Existence and properties of EVA solutions}

Almost identical arguments relating to the soliton existence domain can be
made for the EVA solutions. We again simplify the analysis by rewriting Eqs.
(\ref{eqn:p})-(\ref{eqn:P}) in the following form (for $D=3$)
\begin{align}
p & =\frac{q}{2}\left( \frac{5q^{2}+3\gamma}{q^{2}+3\gamma}\right),
\label{eqn:pnew} \\
\alpha_{11} & =\frac{2^{10}p^{3}q^{3}\left[ 4p^{3}-q(p^{2}+3)\right] }{
(2p+q)^{7}(q^{2}+\gamma)(p^{2}-3)},  \label{eqn:qnew} \\
Q & =\frac{(2p+q)^{4}(p^{2}-3)}{3\cdot2^{3}p^{3}(q-2p)},  \label{eqn:Qnew} \\
P^{2} & =\frac{(2p+q)^{3}(q^{2}+\gamma)}{2^{3}q^{3}}Q.  \label{eqn:Pnew}
\end{align}
Here, the expression (\ref{eqn:qnew}) for $\alpha_{11}$ is obtained from Eq. (\ref
{eqn:q}) and has been further substituted into Eq. (\ref{eqn:Q}) to obtain
Eq. (\ref{eqn:Qnew}). The parameter $p$ in Eqs. (\ref{eqn:qnew})-(\ref
{eqn:Pnew}) is to be substituted using Eq. (\ref{eqn:pnew}). By treating $q$
as a free parameter instead of $\alpha_{11}$ (assuming $q>0$ for localized
solutions), we can first solve for $p$ in terms of two independent
parameters $q$ and $\gamma$, and then proceed to find the remaining
parameters, $\alpha_{11}$, $Q$, and $P$.

\begin{figure}[ptb]
\includegraphics[height=3.9cm]{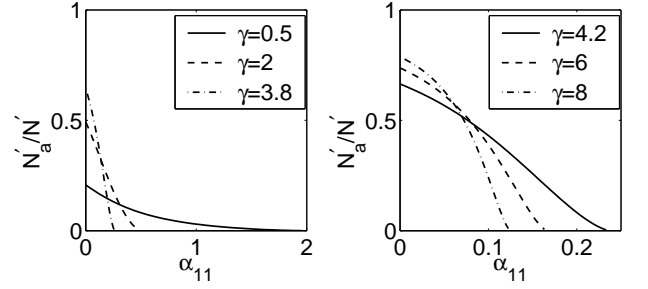}
\caption{Fraction of the number of particles in the atomic field $\mathcal{N}
_{a}^{\prime}/\mathcal{N}^{\prime}$ for the EVA solutions as a function of $
\protect\alpha_{11}$, along the lines of fixed $\protect\gamma$-values as
shown in Fig. \protect\ref{fig:GVA_domain} .}
\label{fig:EVA_fraction}
\end{figure}

In Appendix B, we analyze the above set of
algebraic equations and conclude that for $\gamma>0$ the existence domain
for the EVA solutions is identical to that of the GVA solutions, and is
given by $0\leq\alpha_{11}<1/\gamma$.

Figure \ref{fig:EVA_fraction} shows the dependence of the atomic
number fraction $\mathcal{N}_{a}^{\prime }/ \mathcal{N}^{\prime}$
for the EVA solution as a function of $\alpha_{11}$, for various
$\gamma$. As we see, the salient features of these curves, as well
as the behavior of the EVA soliton widths, are similar to those of
the GVA solutions discussed in the previous subsection.

\section{Dynamical stability}

In order for these variational solutions to prove useful, it is
necessary to identify some correlation between their dynamical
behavior and the existence of actual (exact) stable soliton
solutions. To this end, we have identified the exact stationary
solutions numerically, by means of the numerical relaxation
method, and have checked their stability under dynamical evolution
for a large number of test points in the ($\alpha_{11}$-$\gamma$)
parameter space within the GVA/EVA solutions existence domain,
$0\leq\alpha_{11}<1/\gamma$ for $\gamma>0$. What follows is an
account of the dynamical behavior of the variational
approximations, together with comparisons between this behavior
and the existence of stable stationary points.

\subsection{Stability of the GVA solutions}

\begin{figure}[ptb]
\includegraphics[height=5.4cm]{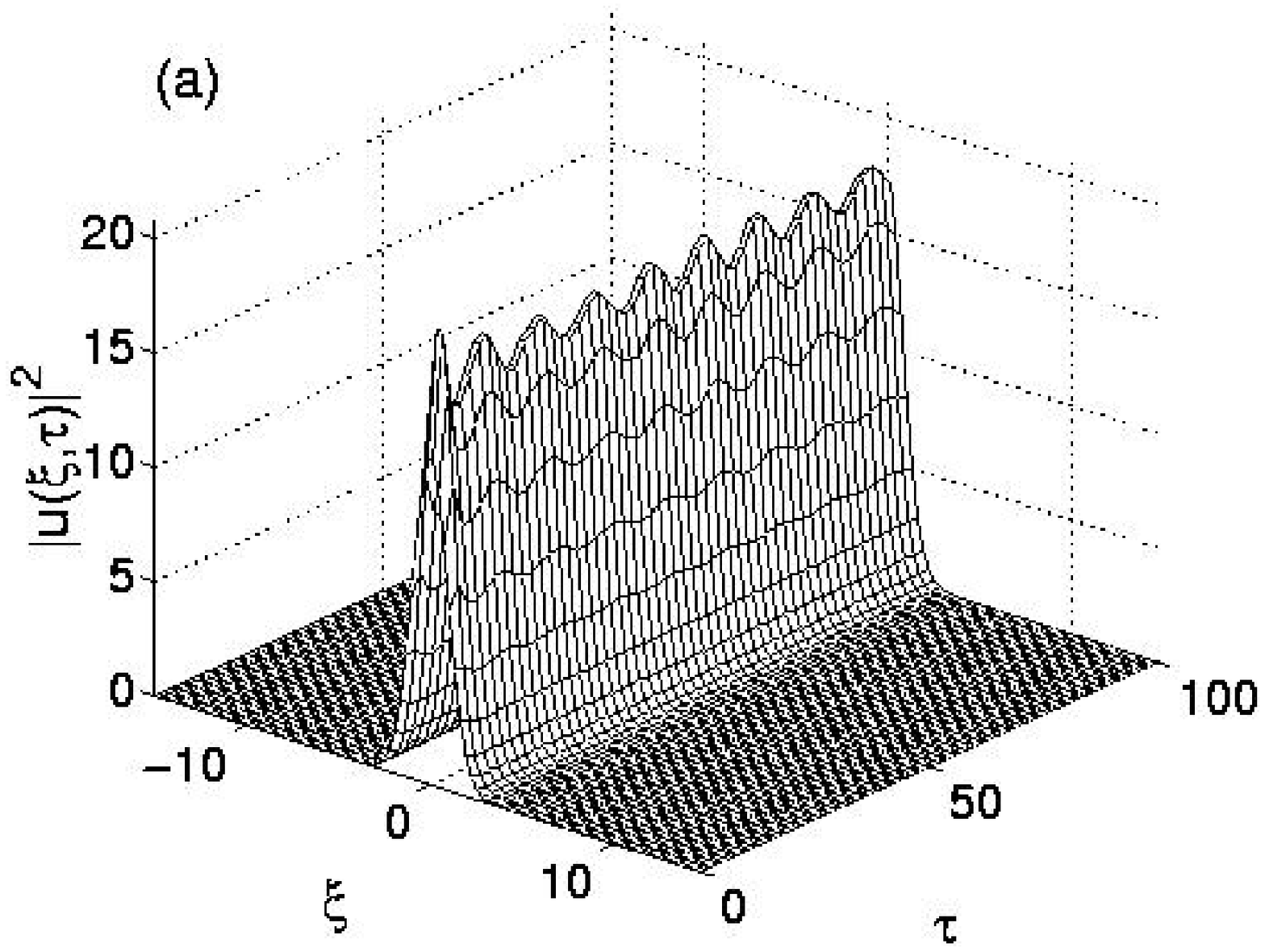}
\par
\includegraphics[height=5.4cm]{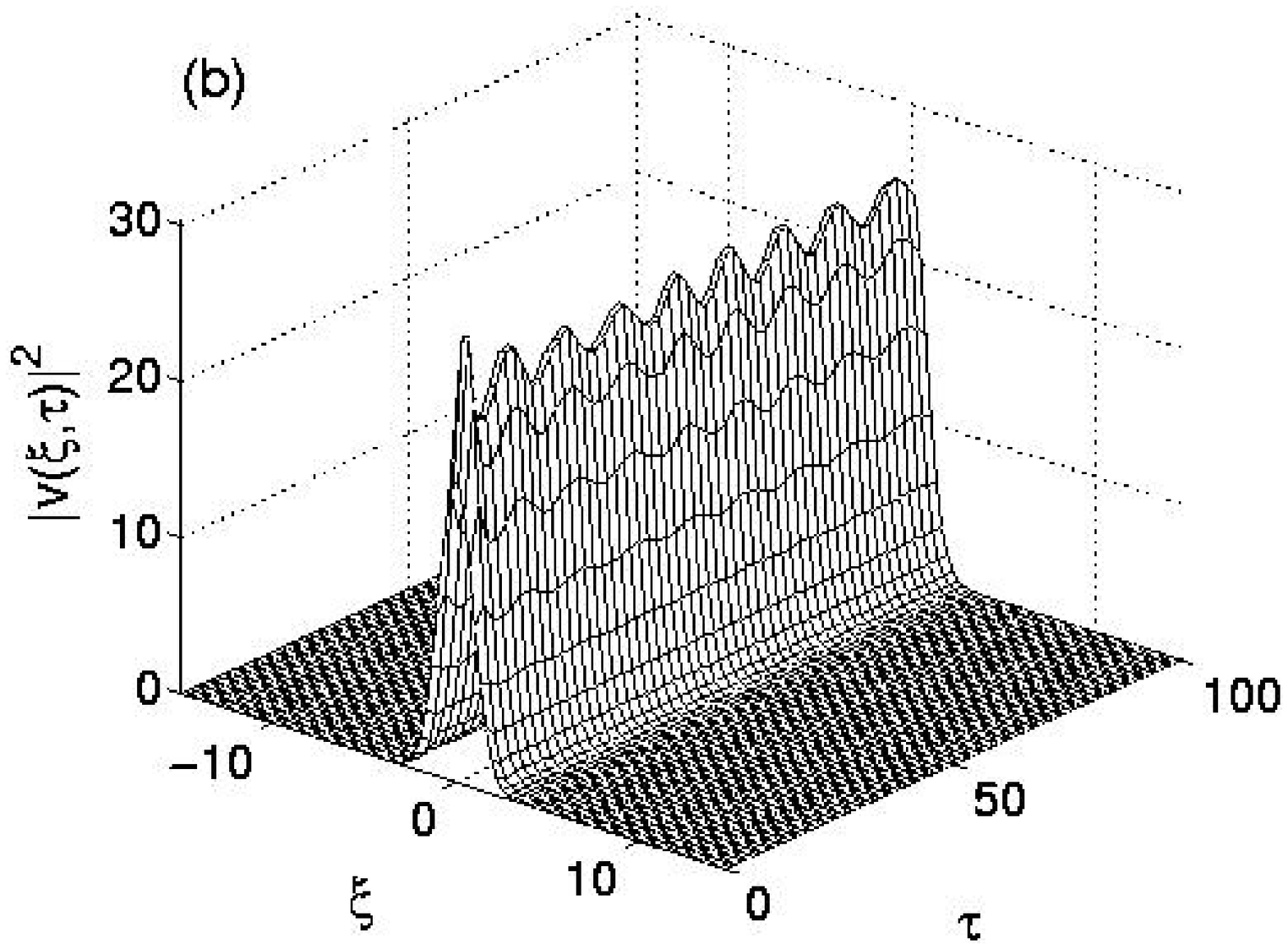}
\caption{Example of stable dynamical evolution of the GVA soliton. Shown are
the particle number densities for the atomic (a) and molecular (b) fields
for $\protect\alpha_{11}=0.1$ and $\protect\gamma=1$, with the GVA solution
taken as the initial condition.}
\label{fig:dynamicsGVA_s}
\end{figure}

\begin{figure}[ptb]
\includegraphics[height=5.3cm]{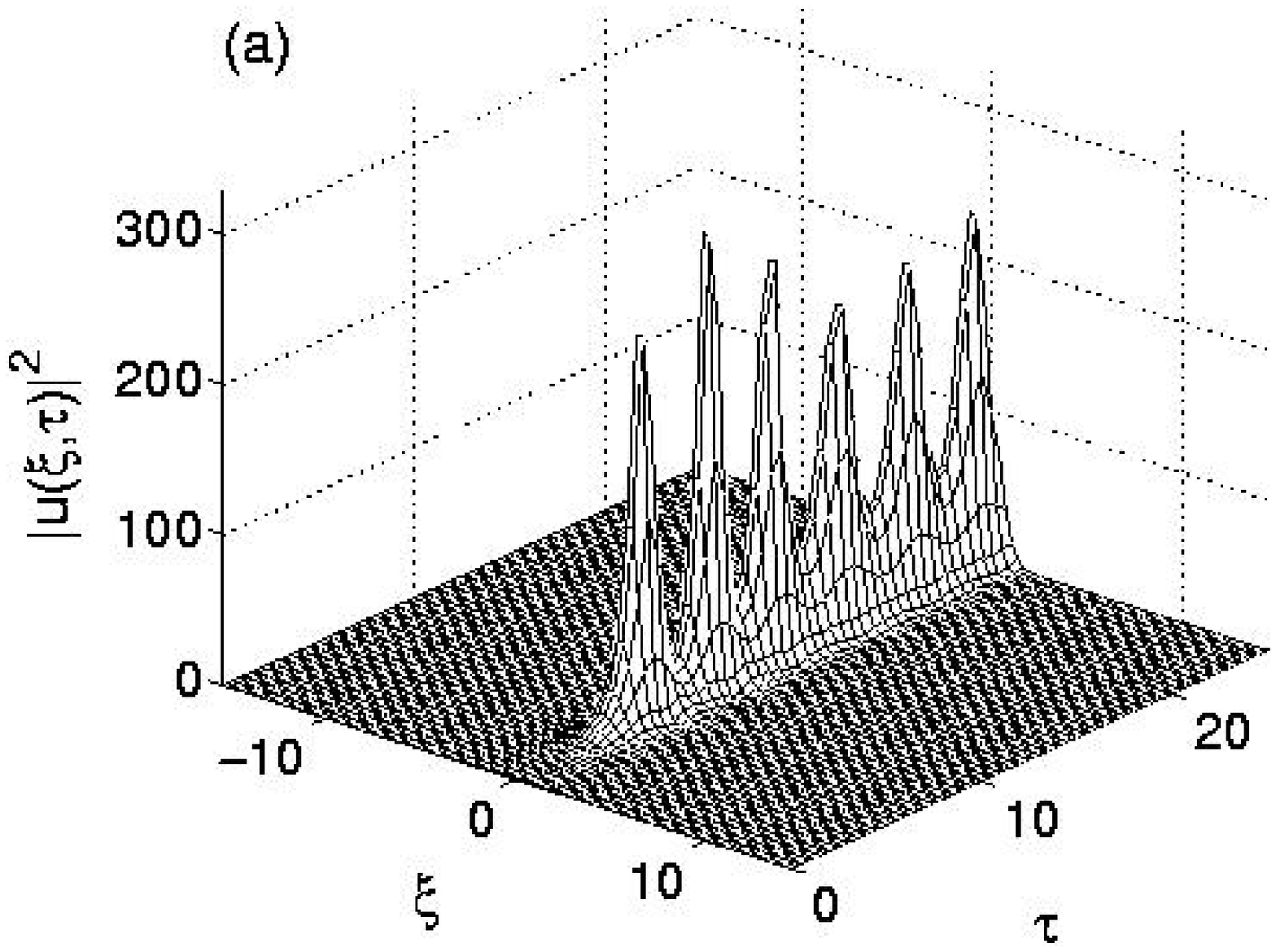}
\par
\includegraphics[height=5.3cm]{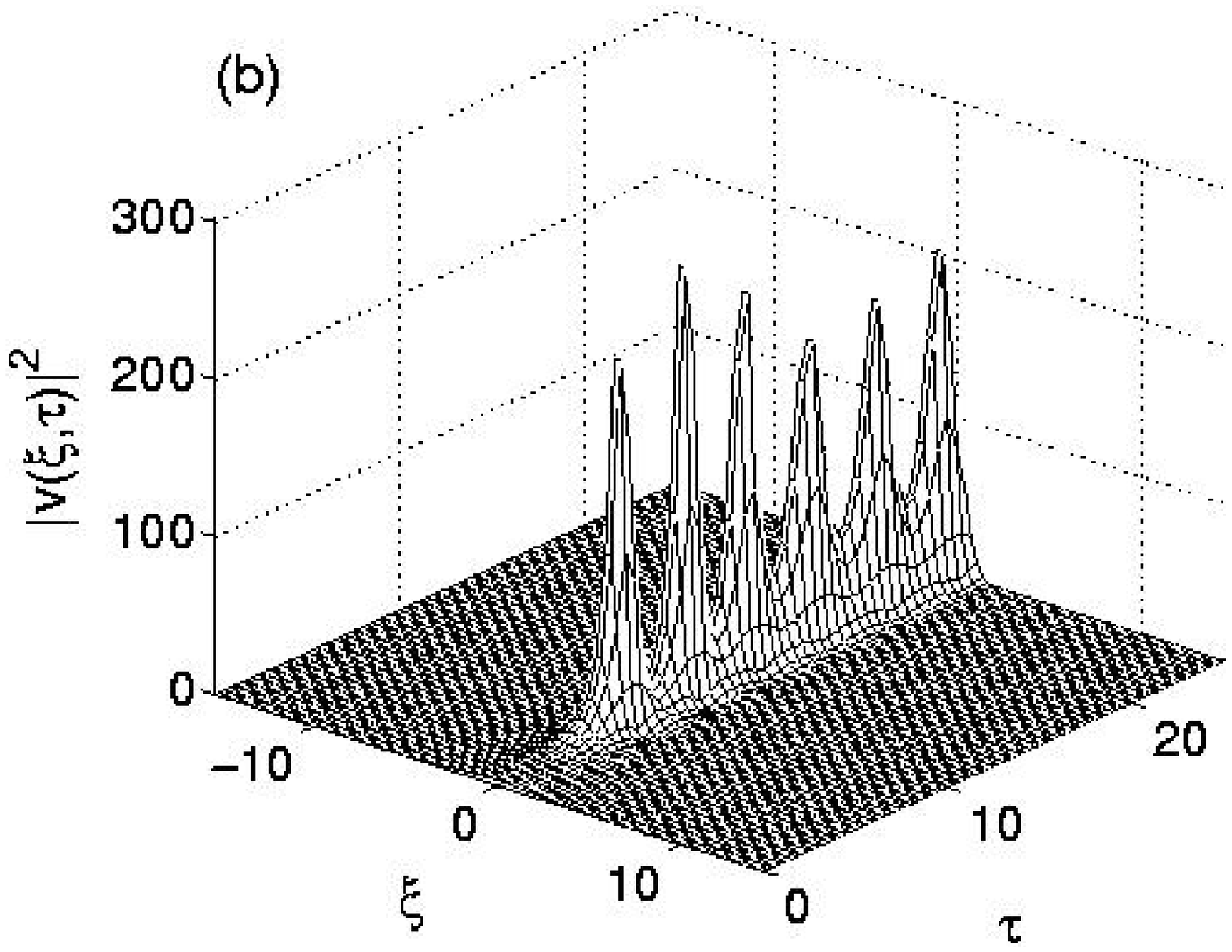}
\caption{Dynamical evolution of the atomic (a) and molecular (b)
field densities for $\protect\alpha_{11}=0.01$ and
$\protect\gamma=0.01$. This is an example representing strongly
`oscillatory' behavior of the GVA solution.}
\label{fig:dynamicsGVA_o}
\end{figure}

\begin{figure}[ptb]
\includegraphics[height=5.3cm]{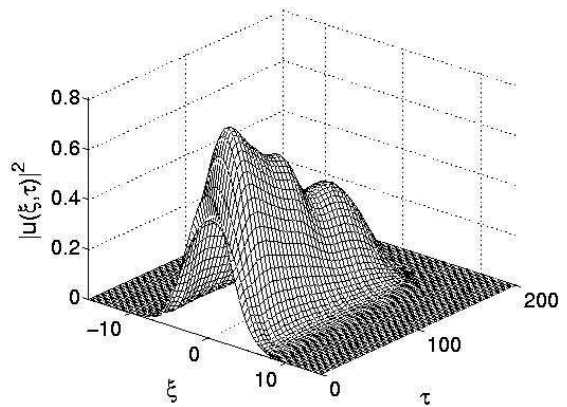}
\caption{Dynamical evolution of the GVA solution for
$\protect\alpha_{11}=3$ and $\protect\gamma=0.01$ representing an
example of unstable behavior. Shown is the particle number density
for the atomic field, with similar behavior observed for the
molecular field.} \label{fig:dynamicsGVA_u}
\end{figure}

We have conducted a numerical analysis of the dynamics of the GVA
solution for various $(\alpha_{11},\gamma)$ pairs lying within the
existence domain $0\leq\alpha_{11}<1/\gamma$, for $\gamma>0$. The
details of this analysis are as follows.

We first used Eqs. (\ref{a-3D})-(\ref{A-3D}) to obtain the
parameters characterizing the GVA solution for each
$(\alpha_{11},\gamma)$ pair in question. These solutions were then
used, in conjunction with the dimensionless equations of motion
(\ref{eqn:nodim-eom}), to form a set of initial value problems.
The dynamical behavior of each Gaussian solution was then
determined through numerical integration using a spherically
symmetric \cite{Comment-spherical-symmetry} semi-implicit
algorithm.

In Figures \ref{fig:dynamicsGVA_s} through \ref{fig:dynamicsGVA_u} we
show typical examples of the dynamical evolution of the atomic and
molecular fields.

\begin{figure}[ptb]
\includegraphics[width=7cm]{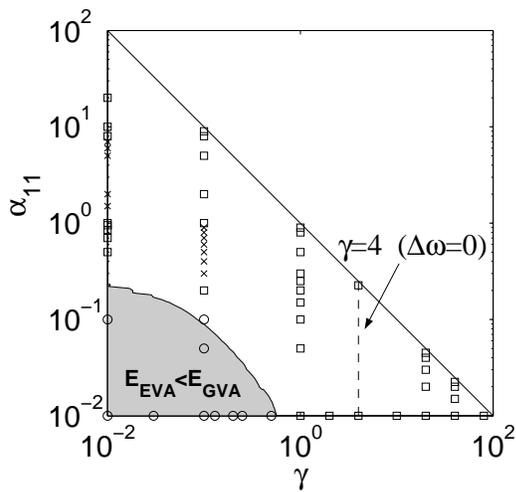}
\caption{Summary of the dynamical behavior of the GVA solutions
for different ($\alpha_{11},\gamma$) pairs. The squares, circles
and crosses indicate stable, strongly oscillatory and unstable
behavior, respectively. For discussion of the shaded region with
$E_{EVA}<E_{GVA}$, see text in Sec. V C.} \label{fig:paper_map1}
\end{figure}

Figure \ref{fig:paper_map1} summarizes the results of our
dynamical stability analysis, applied to many
$(\alpha_{11},\gamma)$ pairs satisfying $\alpha_{11}\geq0.01$, $
\gamma\geq0.01$ and $\alpha_{11}<1/\gamma$. Here, the points
marked with squares, circles or stars represent dynamics of the
GVA solutions, which have been classified as stable,
`oscillatory', or unstable in nature. (This necessarily involves a
certain degree of ambiguity when distinguishing between the stable
and oscillatory cases.) Here, the term oscillatory
is used in a broad sense, and does not mean to imply true periodic
oscillations around the original GVA solution or the exact
stationary solution. The term unstable, on the other hand, refers
to delocalization of the GVA solution over short time-scales
\cite{Comment-time-scales}.

Remarkably, the GVA solutions display primarily stable dynamics
within the $0\leq\alpha_{11}<1/\gamma$ parameter space. Exceptions
to this are two regions close to the $\gamma=0.01$
axis. The lower of these regions (the shaded region in Fig.
\ref{fig:paper_map1} containing circles) contains GVA solutions
which, although remaining localized, display highly oscillatory
dynamics. Here, the EVA solutions have lower energy and give a
better approximation to the exact solutions (see Sec.~V~C for
further discussion). The other region contains unstable GVA
solutions (marked by crosses) which delocalize rapidly under
dynamical evolution.

We point out, however, that these regions are rather small in the
physical parameter space (notice the logarithmic scale in Fig. \ref
{fig:paper_map1}). The most interesting area in this sense is about the $
\gamma=4$ axis, corresponding to a minimum energy mismatch between the
atomic and molecular fields, $\hbar\Delta\omega\simeq0$. For example, even
for the detunings as large as $\Delta\omega\simeq-10^{4}$ s$^{-1}$, which
gives an energy mismatch comparable in magnitude with typical mean-field
energies in atomic Bose-Einstein condensates, the corresponding value of
$\gamma$ is of the order of $\gamma\simeq2$, for typical values of
$N\simeq10^{5}$ and $\chi \simeq10^{-6}$ m$^{3/2}$/s (see also Sec. VI). In
this physically interesting region, the GVA solutions are a good
approximation to the exact solitons and maintain excellent dynamical
stability.

\subsection{Existence of numerical exact solutions}

By using the approximate GVA solution as an initial guess in
the numerical relaxation algorithm, we have investigated the
shape of the exact stationary solution having the same
particle number as the GVA. This constraint is used to ensure that
the numerically
found exact solution corresponds to the same set of physical
parameters as the GVA. The stability of each exact solution
was determined in the same manner as that of the GVA,
i.e. via real-time dynamical evolution governed by Eqs.
(\ref{eqn:nodim-eom}). In all cases where a stationary solution was
identified but found to be unstable, a modified
initial guess was found that converged to a \emph{stable} stationary
solution. The modified Gaussian used in these cases was typically
narrower and of higher peak density, while having the same total
particle number. In a small subset of cases, no exact stationary
solution was obtained (see below).

\begin{figure}[ptb]
\includegraphics[height=5.3cm]{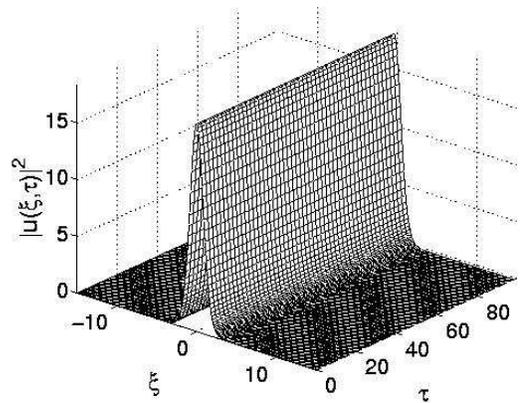}
\caption{Exact soliton dynamics with the exact solitary solution
(found numerically) as the initial condition, for
$\protect\alpha_{11}=0.1$, $\protect \gamma=1$, and
$\mathcal{N}^{\prime}$ being fixed to the same value as in the
respective GVA solution. Shown is the particle number density for
the atomic field, with similar behavior found for the molecular
field.} \label{fig:exact_propagation}
\end{figure}

Figure \ref{fig:exact_propagation} illustrates the time evolution of one
such solution. This demonstrates the stability of the exact soliton
solution, in contrast to those corresponding to energy maxima which become
delocalized due to the buildup of small numerical inaccuracies such as
rounding errors.

\begin{figure}[ptb]
\includegraphics[width=7cm]{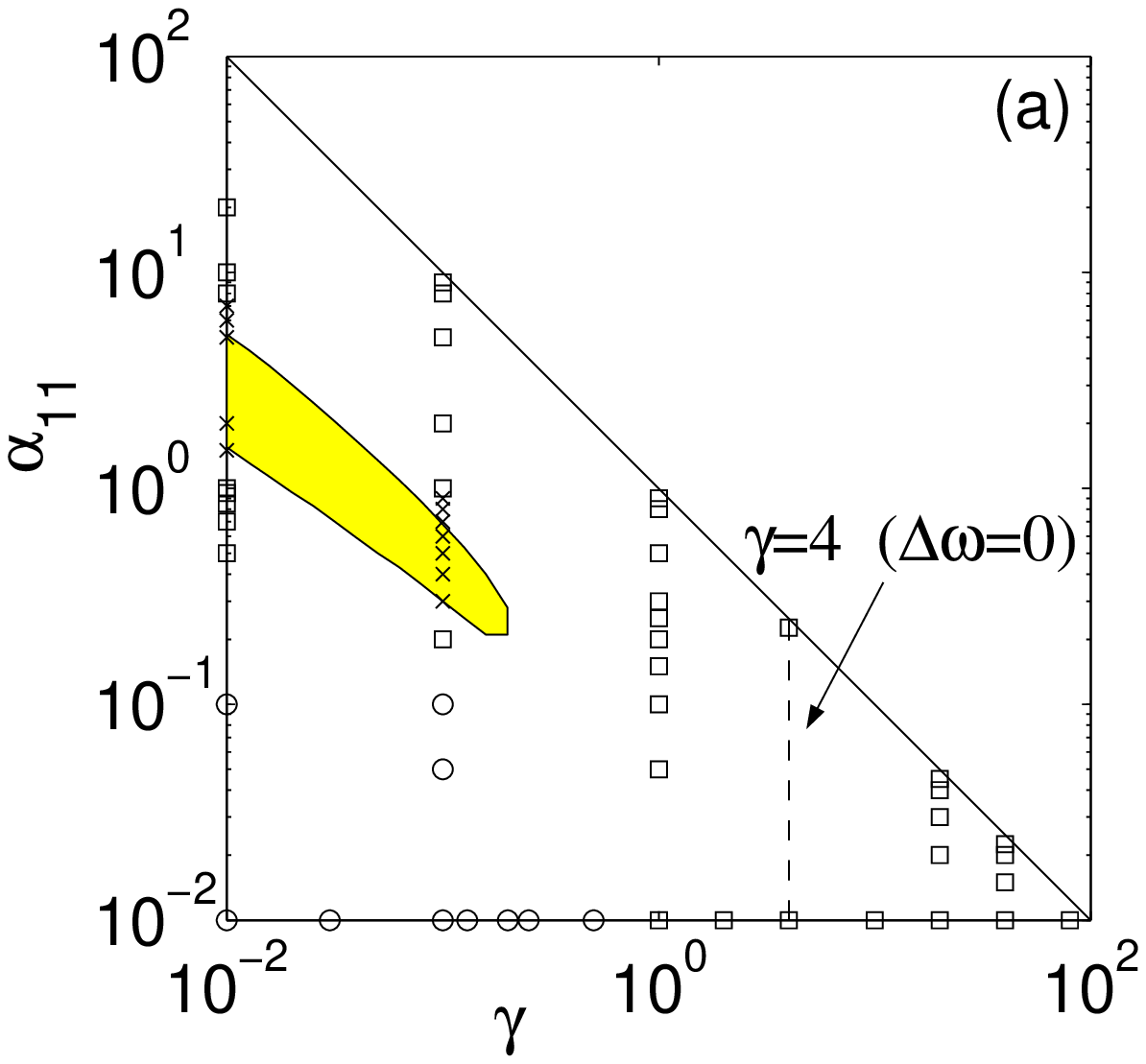}
\par
\includegraphics[width=6.1cm]{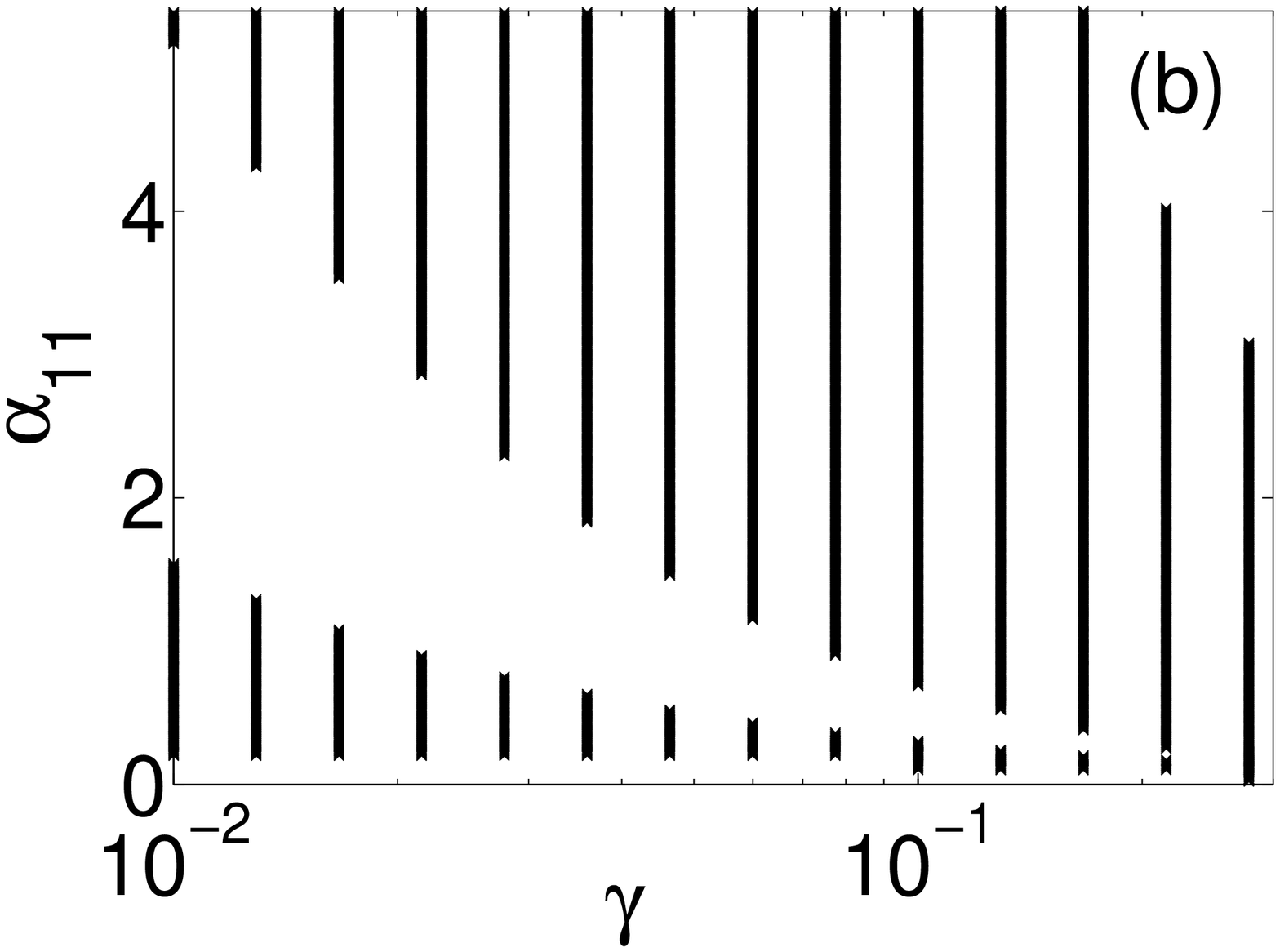}
\caption{(a) Comparison of stable soliton existence domain with
the dynamical behavior of GVA solutions. Stable exact solutions
were found for all marked points outside of the shaded region. The
precise boundary of the shaded region, where no exact solutions
were found, was identified by approaching it from below and from
above along the vertical lines shown in (b). The lines themselves
consist of points, of spacing $0.01$ in $\alpha_{11}$, for which
the exact solutions were identified, while the interrupted part of
each line corresponds to having no exact solutions.}
\label{fig:paper_map2}
\end{figure}

In Fig.~\ref{fig:paper_map2}, we summarize the results of the
stability analysis for the exact solutions and compare them with
those of the GVA. Stable exact soliton solutions were
identified for all $(\alpha_{11},\gamma)$ pairs used in the GVA
dynamical analysis (see Fig. \ref{fig:paper_map1}), except those
lying within the shaded region. The boundary of this region was
found using an extra set of test points for higher accuracy.
Distorting the initial guess input to the
relaxation algorithm did not help in finding exact stable
solutions in the shaded region. Note that, apart from a small
number of exceptions near the boundary, all
unstable GVA solutions in Fig.~\ref{fig:paper_map1} which
delocalized under dynamical evolution (as for the example in
Fig.~\ref{fig:dynamicsGVA_u}) are contained within the shaded
region in Fig.~\ref{fig:paper_map2}. The physical origin of this
instability is yet to be understood.

Thus, one can make the statement that the existence \emph{and}
dynamical stability of the GVA solution is strongly indicative of
the existence of a true stationary soliton solution. This applies
even to the case of strongly oscillatory GVA dynamics, in the
sense that we were able to find a stable exact soliton solution
whenever the oscillatory behavior of the GVA persisted for long
evolution time.

\subsection{Oscillatory dynamics and EVA solutions}

Stationary solitons for ($\alpha _{11},\gamma $) pairs in the lower left
corner of Figure \ref{fig:paper_map1} are poorly approximated by the Gaussian
ansatz solutions -- a fact revealed by the oscillatory GVA dynamics
prevalent in this area. Examining the profiles of the corresponding
numerically-obtained exact solutions suggests that the EVA solutions may
provide a better approximation in this region.

\begin{figure}[ptb]
\includegraphics[height=4.7cm]{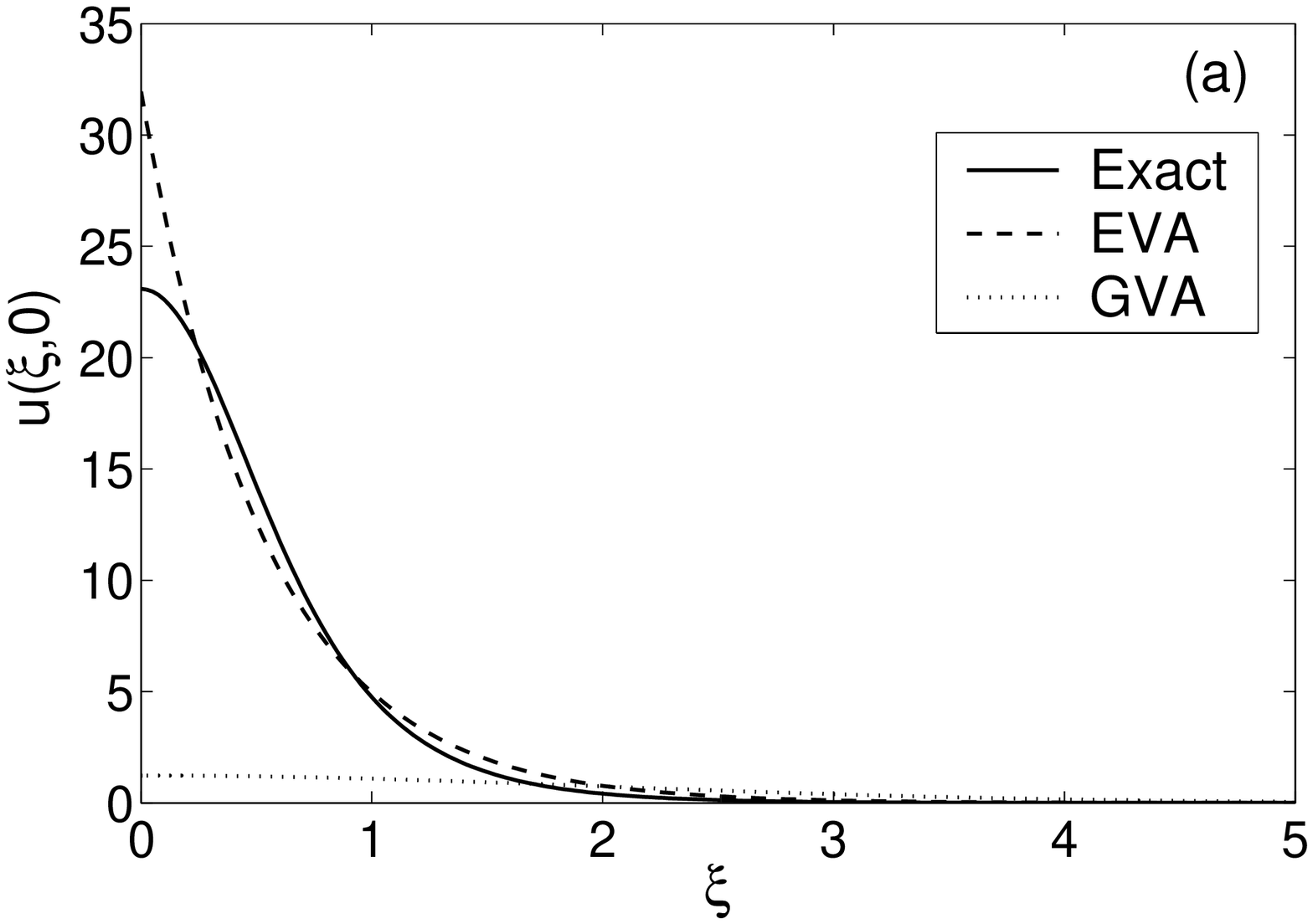}
\par
\includegraphics[height=4.7cm]{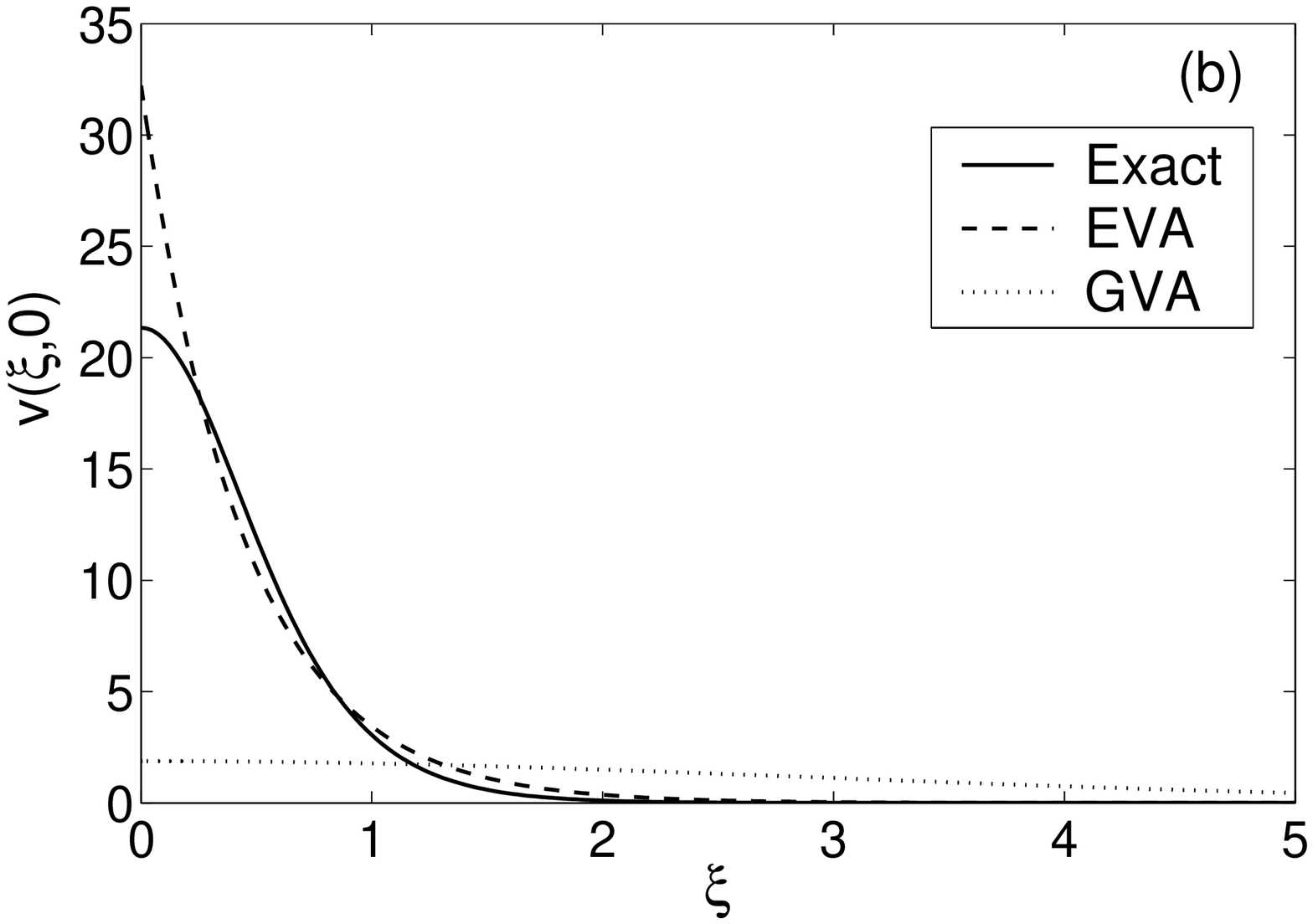}
\caption{Comparison of the density profiles for the atomic (a) and the
molecular (b) fields described by the GVA, EVA and the exact stationary
solutions, for $\protect\alpha_{11}=0.01$ and $\protect\gamma=0.01$. Note
the dramatic failure of the GVA to approximate the exact solution in this
case.}
\label{fig:eva_compare}
\end{figure}
\begin{figure}[ptb]
\includegraphics[height=5.3cm]{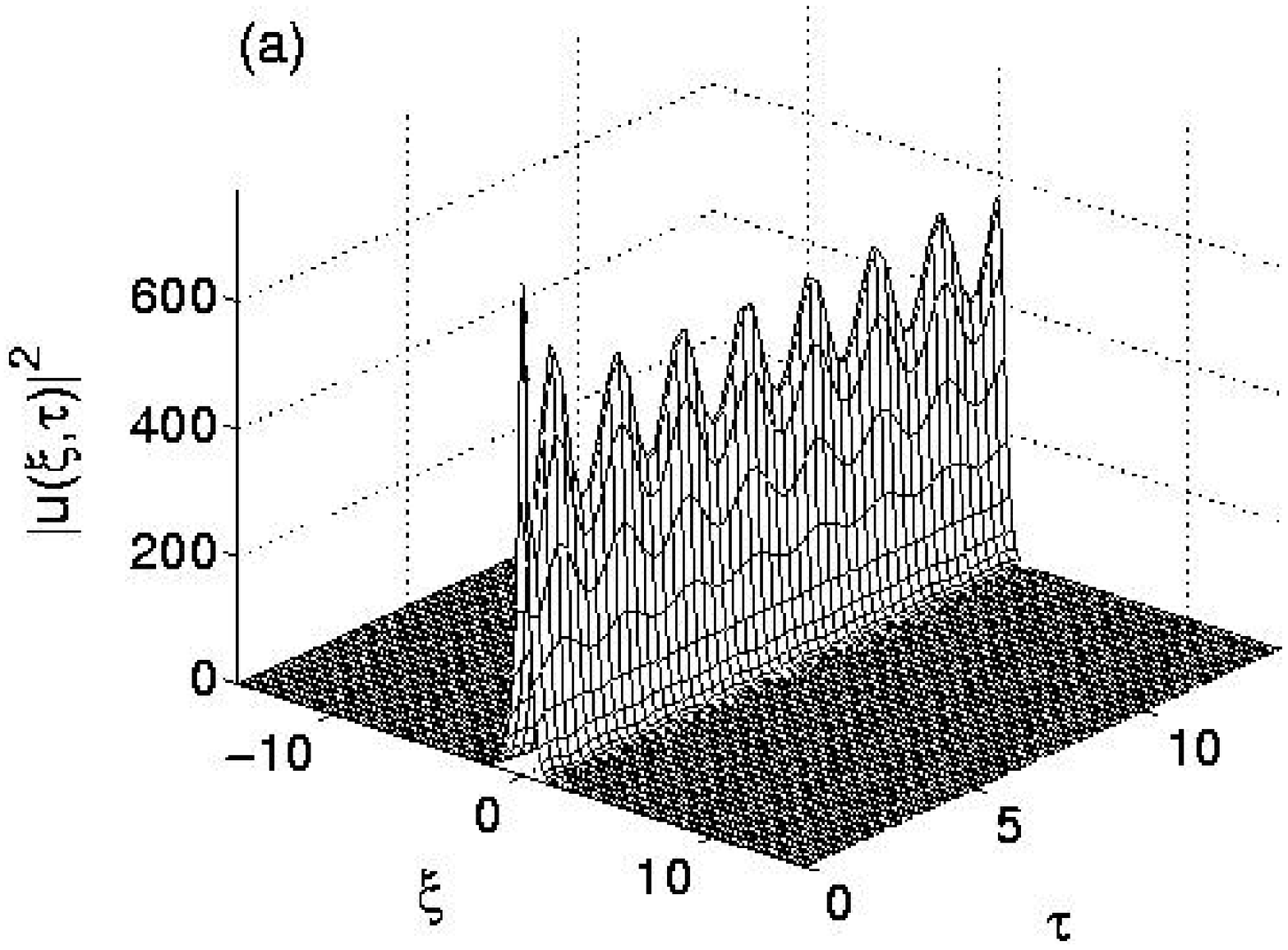}
\par
\includegraphics[height=5.3cm]{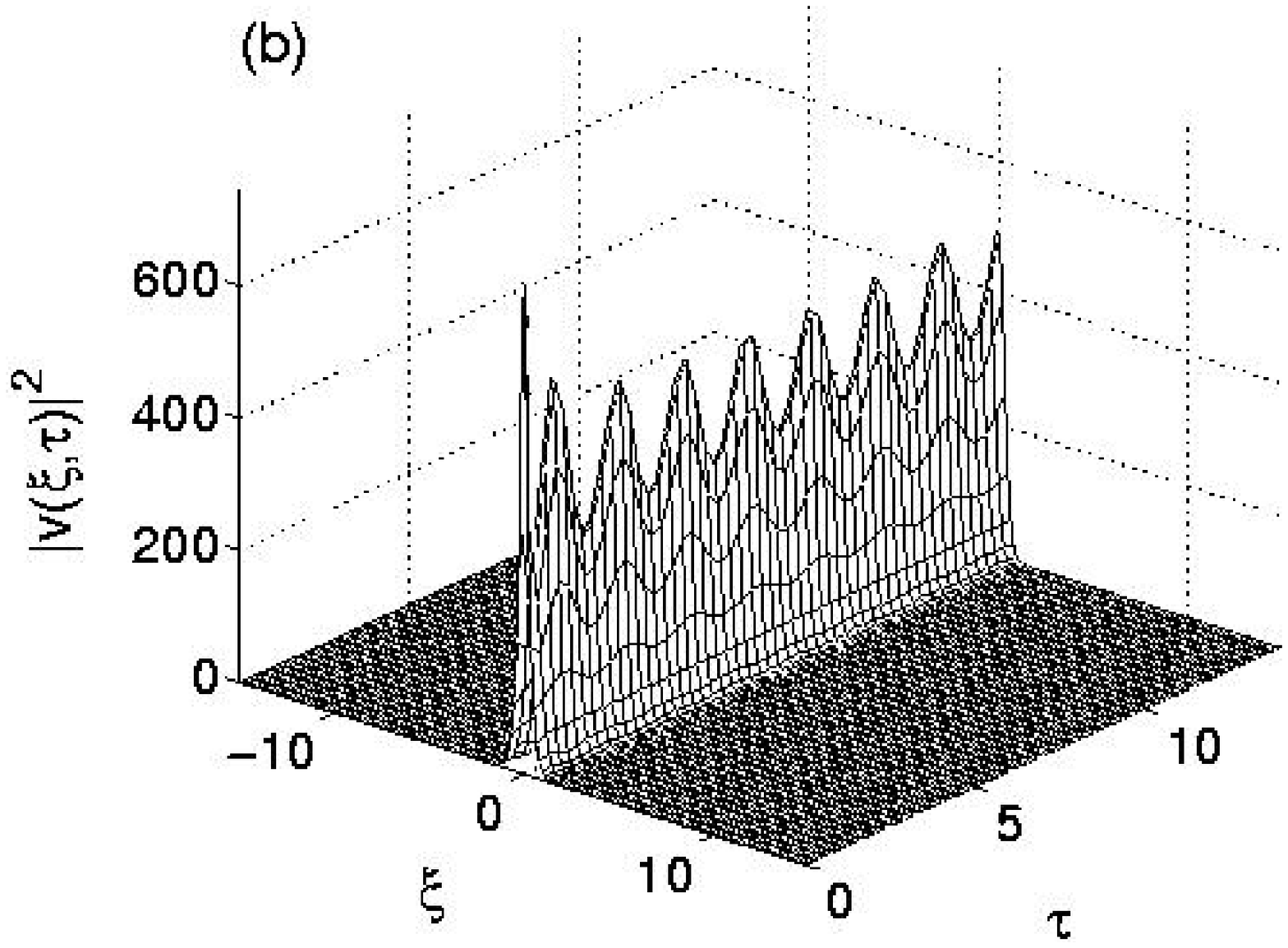}
\caption{Dynamical evolution of the EVA solution for $\protect\alpha
_{11}=0.01$ and $\protect\gamma=0.01$. Shown are the particle number
densities for the atomic (a) and molecular (b) fields. This figure can be
compared with the strongly oscillatory dynamics of the GVA solution in Fig.
\protect\ref{fig:dynamicsGVA_o}, and shows improvement in the stability of
the EVA solution due to its lower energy.}
\label{fig:eva_prop}
\end{figure}

In order to test this for a given ($\alpha_{11},\gamma)$ pair, we need to
ensure that the GVA and EVA solutions in question are being compared for the
same value of the parameter $\mathcal{N}^{\prime}$ corresponding to the
total number of particles. Thus, we first identified the value of $\mathcal{N}
^{\prime}$ for the GVA solution, and then solved
Eqs. (\ref{eqn:constrained_EVA1})-(\ref{eqn:constrained_EVA3}) for
the parameters of the corresponding \emph{constrained} EVA solution.

Figure \ref{fig:eva_compare} illustrates the improvement in the fit of the
profile of the EVA solution to that of the exact stationary solution, for $
\alpha_{11}=\gamma=0.01$. The corresponding GVA solution is also shown for
comparison. The resulting reduced amplitude of oscillation in
the dynamics of the EVA is shown in Fig. \ref{fig:eva_prop}.

In order to understand and quantify this improvement, the total
Hamiltonian energy corresponding to both ansatz solutions has been
calculated for a collection of points spanning the
parameter space under consideration. The shaded area with
$E_{EVA}<E_{GVA}$ in Fig. \ref{fig:paper_map1} represents the
region of parameter space where the constrained
EVA solutions have lower energy than that of the GVA. This
analysis shows that the EVA indeed provides a better approximation
to the exact solitons in cases when the dynamics of the GVA is
strongly oscillatory.

Previous investigations of the SN equation \cite{SN-2},
which has a stationary solution exactly equivalent to ours with $\gamma
=\alpha _{11}=0$, have come to the same conclusion that over a class of
trial functions the exponential ansatz for the `atomic' field provides the
best upper bound to the ground state energy. The fact that
the upper bound provided by our solution,
$-0.108m^{5}G^{2}N^{2}/\hbar ^{2}$,
is higher than the value $-0.146m^{5}G^{2}N^{2}/\hbar ^{2}$
quoted in \cite{SN-2} is due to the fact that we have used variational
solutions for both $u$ and $v$-fields, rather than just for $u$.
These upper bounds can be
contrasted with the exact ground state energy of $-0.163m^{5}G^{2}N^{2}/
\hbar ^{2}$ \cite{SN} which we have verified using our numerical relaxation
code.

The SN helps to also understand the
dramatic failure of the GVA solutions to approximate the exact
solitons in the region of small $\gamma$ and $\alpha$. Here, the
tails of the density distributions become increasingly important,
and according to Eq. (\ref{exp-tails}) the $v$-field with
$\gamma\rightarrow0$ decays much slower than the tail of any
Gaussian. In terms of the SN equation, this corresponds to
evolution of the wave-function in a very long-range potential $v$.

\subsection{Energy surface topography}

In order to better understand the nature of the instability
marked by the shaded region in Fig. \ref{fig:paper_map2}, it
would be instructive to examine the topographical variations in
the energy surface while approaching and crossing the boundary
between the stable and the unstable regions. By tracing a single
dimensional trajectory between known critical points in the
infinite dimensional coordinate space we can learn something of
the higher-dimensional structure.

\begin{figure}[ptb]
\includegraphics[height=5.4cm]{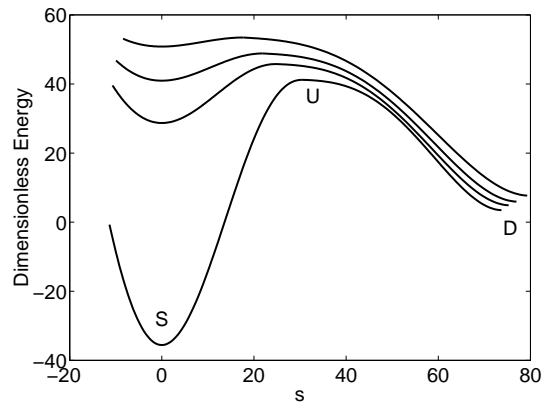}
\caption{Energy surface topography as one approaches the unstable region
boundary from below along the $\protect\gamma=0.01$ line. Shown is the
dimensionless Hamiltonian energy vs the distance $s$ in state
space \cite{Comment-trajectories}.
Different curves correspond (from bottom in increasing order) to
$\protect\alpha_{11}=0.2,0.5,0.7,1$. }
\label{fig:topo}
\end{figure}

Figure \ref{fig:topo} illustrates the topographical variation in
the energy surface when the unstable region is approached from
below as $ \alpha_{11}$ is increased along the $\gamma=0.01$ line.
Here, the critical points (including the unstable or delocalized
solutions) for each ($ \alpha_{11}$,$\gamma$) pair were evaluated,
and linear interpolation was used to generate a continuous path
between stable ($S$), unstable ($U$) and delocalized ($D$)
solutions, along a contour of constant $\mathcal{N}^{\prime}$
\cite{Comment-trajectories}.

It is clear from this illustration that, as $\alpha_{11}$ approaches the
unstable region, the soliton solution becomes metastable as it corresponds
to a local rather than absolute minimum in the energy surface. This fact
does not affect the existence of stable solutions, however, which persist
until the local minimum is completely eliminated as the boundary is crossed
into the unstable region.

We can use this analysis to explain the existence of GVA solutions
which delocalize under evolution despite the existence of an exact
stationary solution. In such cases, the width of the confining
local minimum well (in state space) is smaller than the
perturbation to the exact solution induced by enforcing the
Gaussian ansatz. As one might expect, this only happens near a
stability boundary, where the wells are small.


\section{Relation to physical parameters}

From the practical point of view, the question of interest is
whether a given set of the physical parameters $m_{1}$, $\chi $,
$\kappa _{11}$, $\Delta \omega $, and the total number of
particles $N$ can support stable atomic-molecular solitons. In
order to answer this question we must be able to inter-convert
between the soliton parameters in terms of the dimensionless
variables and the original physical parameters. This procedure is
not of a trivial matter, and requires a self-consistent solution
that is able to map a given set of values of ($m_{1}$, $\chi $,
$\kappa _{11}$, $ \Delta \omega $, $N$) into a pair of values of
($\gamma ,\alpha _{11}$), using the time scale $t_{0}$ and the
length scale $d_{0}$. Depending on whether or not the pair of
values of ($\gamma ,\alpha _{11}$) is inside the soliton existence
domain $\gamma >0$ and $0\leq \alpha _{11}<1/\gamma $, one can
then answer the above question and find the soliton parameter
values in terms of dimensional variables, using the results of
previous sections.

We recall that the relationship between the parameters of the dimensionless
system and the original physical parameters are as follows:
\begin{equation}
\Delta\omega =(\gamma-4)/2t_{0},  \label{eqn:convert1}
\end{equation}
\begin{equation}
\kappa_{11}/\chi^{2} =\alpha_{11}t_{0},  \label{eqn:convert2}
\end{equation}
\begin{equation}
N=\frac{1}{\chi^{2}\sqrt{t_{0}}}\left( \frac{\hbar}{2m_{1}}\right) ^{D/2}
\mathcal{N}^{\prime}.  \label{eqn:convert3}
\end{equation}

Here, the role of the mass $m_{1}$ is in setting up the length
scale $d_{0}= \sqrt{\hbar t_{0}/2m_{1}}$, so that the soliton
widths are found unambiguously once the corresponding time scale
$t_{0}$ is set up self-consistently. This leaves us, in general,
with four independent variables ($\chi$, $\kappa_{11}$,
$\Delta\omega$, $N$) in the physical parameter space and two
independent variables ($\gamma,\alpha_{11}$) of the dimensionless
system. Therefore, in order to be able to map the soliton
existence domain in the ($\gamma,\alpha_{11}$) plane into the
physical parameter space, we have to restrict ourselves to cases
where -- out of four physical parameters ($\chi$, $\kappa_{11}$,
$\Delta\omega$, $N$) -- only two can be varied independently,
while the other pair must be kept fixed.

\begin{figure}[ptb]
\includegraphics[height=4cm]{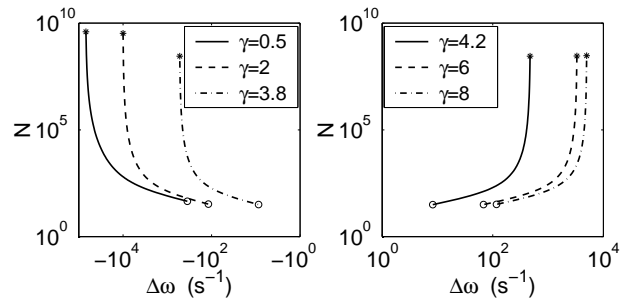}
\caption{Variation in $N$ and $\Delta\protect\omega$ along the ($\protect
\alpha_{11},\protect\gamma$) lines shown in Fig. \protect\ref{fig:GVA_domain}.
The values of the couplings $\protect\chi$ and $\protect\kappa_{11}$ are
held constant: $\protect\chi=10^{-6}$ m$^{3/2}/$s and $\protect\kappa
=4.96\times10^{-17}$ m$^{3}/$s.}
\label{fig:N-deltaomega}
\end{figure}

Depending on which pair of the physical parameters is chosen to be fixed or
varied, one can identify six different cases where the solution of the
problem can be found unambiguously. As an example we consider the case where
the fixed pair of the parameters are the couplings $\chi$ and $
\kappa_{11}$, while the adjustable parameters are the detuning $\Delta\omega$
and the total number of particles $N$. This is the most physically relevant
case, as the detuning and the total number of particles are easier to vary
experimentally.

The procedure of mapping the soliton existence domain in the ($\gamma
,\alpha_{11}$) plane into the ($\Delta\omega,N$) parameter space consists of
solving Eqs. (\ref{eqn:convert1})-(\ref{eqn:convert3}) to firstly identify $
t_{0}$, using Eq. (\ref{eqn:convert2}), and then finding the respective
values of $\Delta\omega$ and $N$ from the remaining two equations. As an
intermediate step, this involves the evaluation of $\mathcal{N}^{\prime}=
\mathcal{N}^{\prime}(\gamma,\alpha_{11})$ using Eq. (\ref{N-prime(A,B,a,b)}
), where the soliton parameters $a,b,A$ and $B$ are found from the GVA
solutions for the ($\gamma,\alpha_{11}$) pair in question. Figure \ref
{fig:N-deltaomega} demonstrates this mapping for the values of $\kappa_{11}$
and $\chi$ typical of a $^{87}$Rb BEC experiments.

Similar mapping can easily be constructed in other cases, where depending on
the choice of the fixed pair of the physical parameters the sequential order
of solving Eqs. (\ref{eqn:convert1})-(\ref{eqn:convert3}) will vary. In all
cases, the initial step should consist of identifying the value of the
`dummy' parameter $t_{0}$ using one of the equations
(\ref{eqn:convert1})-(\ref{eqn:convert3}), and then eliminating
it in favor of the remaining pair
of the physical parameters in question.

\section{Summary}

To summarize, we have applied both Gaussian and exponential variational
approximations to the problem of identifying $3D$ soliton solutions to
parametrically coupled dilute atomic and molecular Bose condensates, with
atomic self-interaction present. The soliton existence domain has been
investigated, and for $\gamma >0$ found to be defined by $0\leq \alpha
_{11}<1/\gamma $ in both cases.

A detailed numerical study of the dynamical behaviour of the
Gaussian ansatz solutions has revealed that, as a rule of thumb,
localized propagation of GVA solutions indicates the existence of
true (and stable) stationary solutions. Furthermore, in regions
where the GVA propagation is strongly oscillatory, the
corresponding exponential ansatz solutions have been found to
possess a lower energy, and propagate almost without oscillations.
This indicates that systems for which the energy of molecular
formation is large and negative and which have weak atomic
self-interaction possess stationary solutions better approximated
by the exponential rather than Gaussian ansatz. Finally, we have
identified an anomalous instability `island' in the ($\alpha
_{11},\gamma $) parameter space, where the Gaussian solitons were
dynamically unstable \textit{and} where no exact stationary
solutions were numerically found.

The results obtained give the precise conditions under which $3D$ coupled
atomic-molecular BEC solitons can form. This is done in terms of the
dimensionless parameters ($\alpha_{11},\gamma$) that originate from physical
parameters determined by the atom-molecule coupling, atom-atom repulsive $s$
-wave scattering, and the energy detuning between the atomic and molecular
fields. The total number of particles $N$ in the system is incorporated in
the analysis self-consistently, via scaling with respect to the time scale $
t_{0}$ or equivalently with respect to the choice of the energy origin. We
have shown how one can in principle map the physical parameter space into
the parameter space defined by the dimensionless parameters ($
\alpha_{11},\gamma$), and hence identify whether a particular set of
physical parameter values, ($\chi,\kappa_{11},\Delta\omega,N$), lays within
the identified soliton existence and stability domain in the ($
\alpha_{11},\gamma$)-plane.

Beyond the theoretical interest, the realization of this type of coupled BEC
solitons could provide a way to stabilize the otherwise diverging output of
an atom-molecular laser, providing a technique for delivering a localized,
intense and coherent atomic-molecular field to a target or a detector.

The steady-state and variational solutions found here are also applicable to
generalizations of the Schr\"{o}dinger-Newton equation.

\begin{acknowledgments}
The authors gratefully acknowledge the ARC for the support of this
work and thank G. Collecutt for helpful discussions. We also thank
the authors of the XMDS software \cite{xmds} which was used here
for dynamical simulations.
\end{acknowledgments}

\appendix

\section*{APPENDIX A}

Standard algebraic analysis of Eqs.~(\ref{a-3D})-(\ref{A-3D}) for possible
solutions, for a given pair of $\alpha_{11}\geq0$ and $\gamma$, gives the
following results. First of all, it is easy to see that for $b>0$ and $B>0$,
the requirement $a>0$ and $A^{2}>0$ (so that $A^{2}B>0$ too) can be
fulfilled only if $\gamma+b>0$. From this, it also follows that $3b+\gamma>0$,
$5b+\gamma>0$, and $b-2a<0$. Next, one can easily see that $B>0$ if $a-1<0$.
This last inequality can be solved in terms of $b$, giving the result
that $b$ must satisfy $b<b_{1}$, where
\begin{equation}
b_{1}=\frac{1}{10}\left[ (2-\gamma)+\sqrt{(2-\gamma)^{2}+40\gamma}\right]
\label{b1}
\end{equation}
corresponds to the positive valued solution of the quadratic
equation for $b$ which follows from $a-1=0$. Thus, as long as
$0<b<b_{1}$ ($a-1<0$) and $ \gamma+b>0$, we satisfy the
requirements that $a>0$, $B>0$, and $ A^{2}>0$.

\begin{figure}[ptb]
\includegraphics[height=4.5cm]{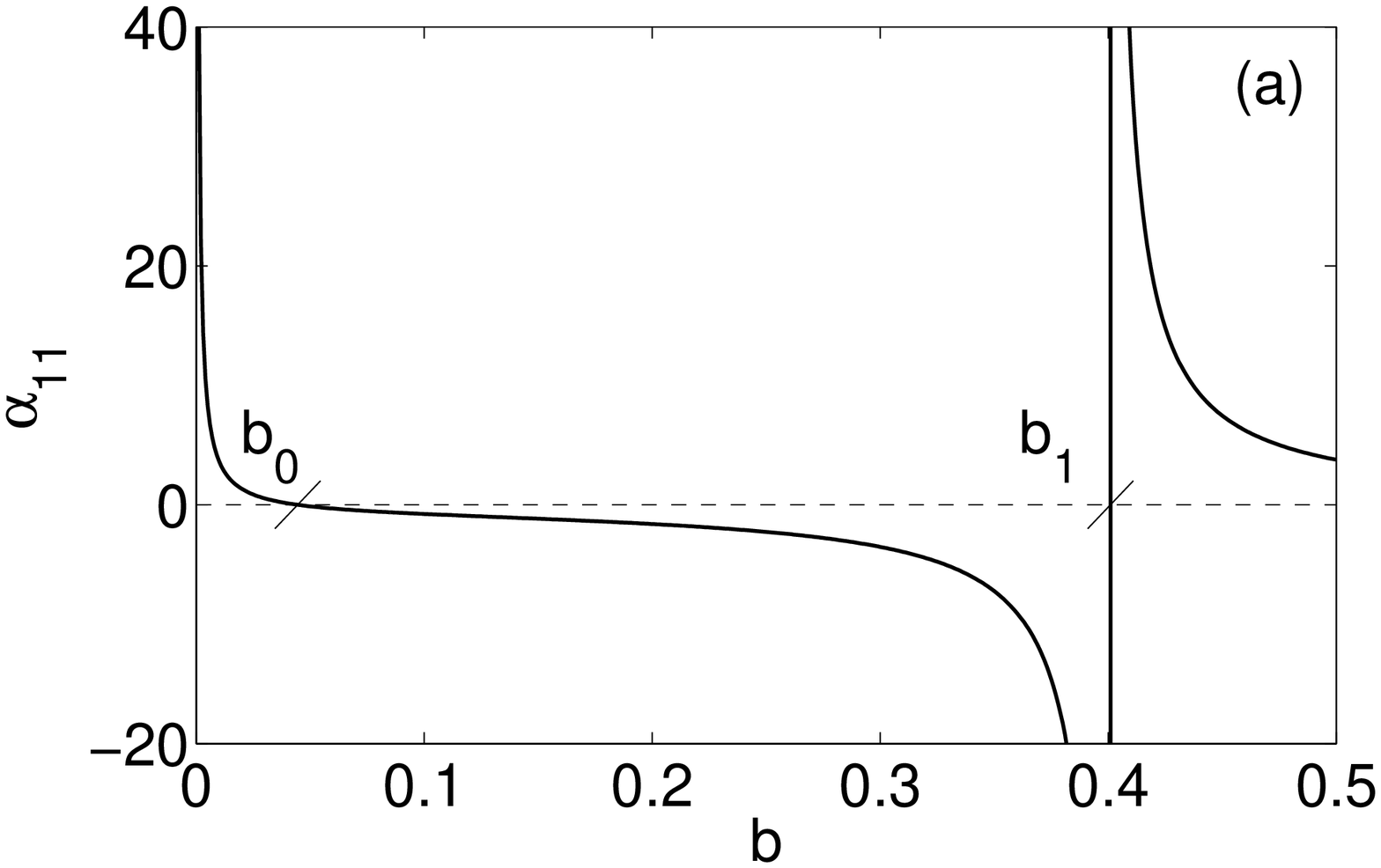}
\par
\includegraphics[height=4.5cm]{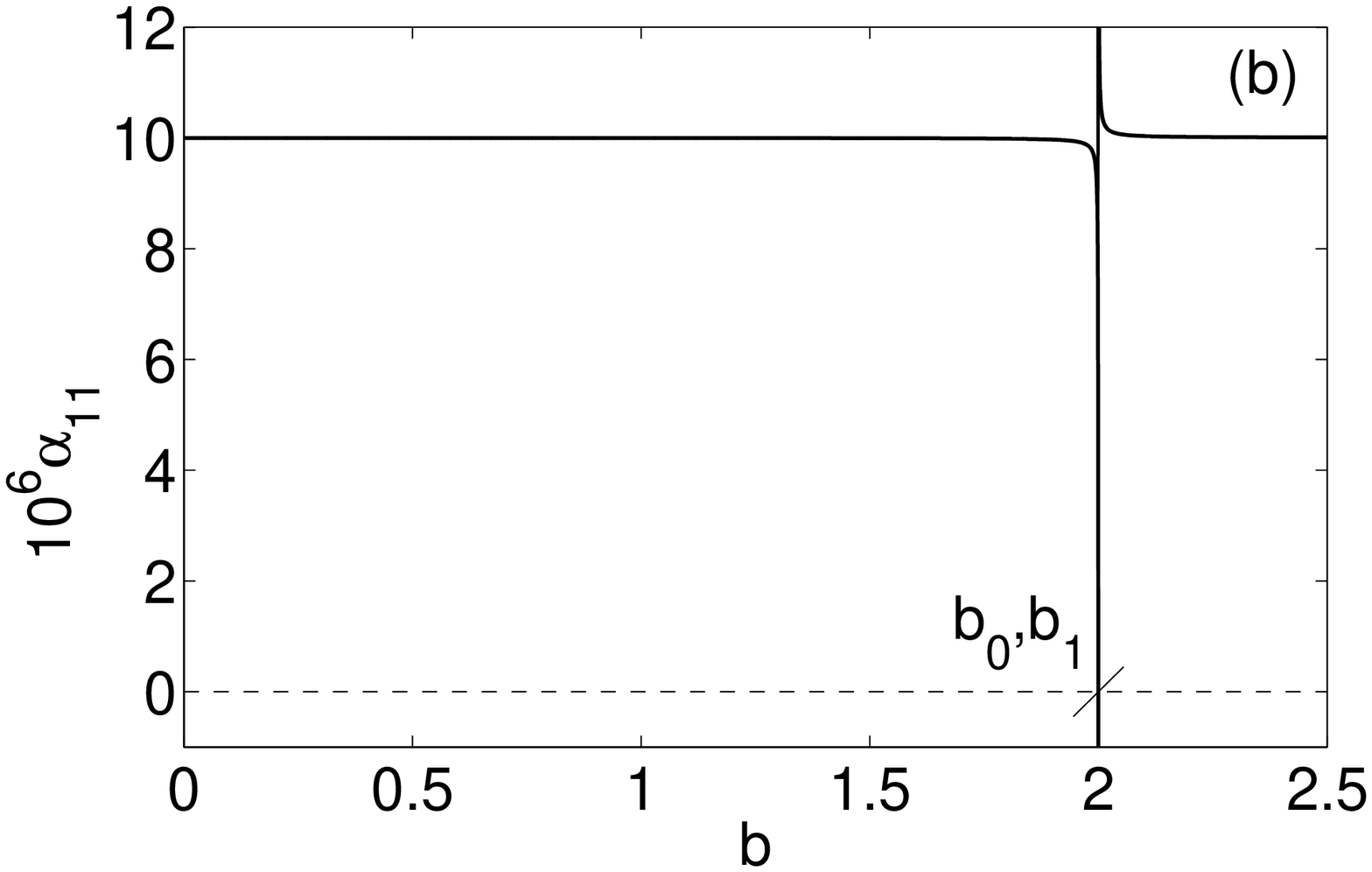}
\caption{The variation in $\protect\alpha_{11}$ as a function of $b$, for
(a) $\protect\gamma=0$ and (b) $\protect\gamma=10^{5}$, illustrating that $
\protect\alpha_{11}$ is monotonically decreasing function of $b$ on $0<b\leq
b_{q}$.}
\label{fig:a11_mondec_gva}
\end{figure}

Analysis of the remaining requirement of $\alpha_{11}\geq0$ is now reduced
to the solution of $4a^{2}-ab-b\leq0$, within the region $0<b<b_{1}$ (where $
a-1<0$) and for $\gamma+b>0$. Here one has to distinguish and consider
separately two cases corresponding to $\gamma\leq0$ and $\gamma>0$.

For $\gamma\leq0$ the analysis is quite complicated and requires a pure
numerical investigation. The overall conclusion is that the soliton
existence domain is now restricted to a very small region which is not of
major physical interest. For example, for $\alpha_{11}=0$, this domain is
limited to the values of $\gamma$ within a narrow interval $
-0.0074<\gamma\leq0$. The size of this interval decreases with increasing $
\alpha_{11}$ and approaches $-1/\alpha_{11}<\gamma<0$ as $
\alpha_{11}\rightarrow\infty$. In terms of the original (physical) phase
mismatch parameter $\Delta\omega$, a physically interesting and important
region corresponds to the values of $\Delta \omega\simeq0$ which we note
correspond to $\gamma\simeq4$, while $\gamma<0$ corresponds to very large
and negative detuning $\Delta\omega$. For this reason, in the remaining of
the paper we only treat the case of $\gamma>0$.

For $\gamma>0$ we proceed with the analytic treatment as follows.
Substituting $a$ from Eq. (\ref{a-3D}), one can rewrite the inequality $
4a^{2}-ab-b\leq0$ (equivalent to $\alpha_{11}\geq0$) in the following form
\begin{equation}
45b^{2}+(14\gamma-2)b+\gamma(\gamma-4)\leq\frac{2\gamma^{2}}{b}.
\label{cubic}
\end{equation}
Taking here the equality sign, and considering the left and the right hand
sides as functions of $b$, gives an equation that can always be solved
graphically, with the result that for $\gamma>0$ there always exists one and
only one real \emph{positive} solution for $b$, which we define via $b_{0}$.
Consequently, the above inequality and therefore $\alpha_{11}\geq0$ is
satisfied if $0<b\leq b_{0}$.

The next step in our analysis consists of showing that $b_{0}<b_{1}$, thus
restricting the soliton existence region to $0<b\leq b_{0}$ from which it
follows that for $\gamma>0$ \ the requirements $a>0$, $\alpha_{11}\geq0$, $
B>0$, and $A^{2}>0$ are satisfied simultaneously. The proof of $b_{0}<b_{1}$
is accomplished by first noting that $b=b_{1}$ (corresponding to $a-1=0$) is
the pole of $\alpha_{11}$, and that $\alpha_{11}$ is a continuous
single-valued function of $b$ within the interval $0<b\leq b_{1}$. The
discontinuity at $b=b_{1}$ is such that $\alpha_{11}\rightarrow-\infty$ as $
b\rightarrow b_{1}-0$ and $\alpha_{11}\rightarrow+\infty$ as $b\rightarrow
b_{1}+0$. In addition, $\lim_{b\rightarrow0}\alpha_{11}=1/\gamma>0$, and
since $b_{0}$ is the only positive root of the cubic equation (\ref{cubic}),
we conclude that the crossover ($b=b_{0}$) of $\alpha_{11}$ from positive to
negative values can only take place at $b<b_{1}$, which implies that $
b_{0}<b_{1}$.

Finally, one can show graphically (see Fig.~\ref{fig:a11_mondec_gva}),
that within the interval $0<b\leq
b_{0}$ and for $\gamma>0$, $\alpha_{11}$ is a monotonically
decreasing function of $b$. It approaches its maximum
$\alpha_{11}\rightarrow1/\gamma$ as $b\rightarrow0$, and
$\alpha_{11}=0$ at $ b=b_{0}$. This implies that there is
one-to-one correspondence between the soliton existence domain
$0<b\leq b_{0}$ and the interval $0\leq\alpha_{11}<1/\gamma$, for
$\gamma>0$. In other words, the interval $0\leq
\alpha_{11}<1/\gamma$ ($\gamma>0$) can be equally regarded
as the soliton existence domain in the parameter space
$(\alpha_{11},\gamma)$.

\section*{APPENDIX B}

Here we analyze the set of algebraic equations (\ref{eqn:pnew} )-(\ref
{eqn:Pnew}) for possible solutions for a given pair of $\alpha_{11} \geq0$
and $\gamma$, with $p$, $q$, $P$ and $Q$ all being real and positive. The
analysis is very similar to the one carried for the GVA solutions, and we
summarize it as follows.

For $q>0$ (as required), in order that $p>0$ it is clearly necessary that $
q^{2}+3\gamma>0$, since in this case we also have $5q^{2}+3\gamma>0$. In
addition, for $P^{2}$ to be positive one has to have $q^{2}+\gamma>0$,
provided that $Q>0$ (as required). The condition $q^{2}+3\gamma>0$ then
leads to the requirement that $p^{2}-3<0$ for $Q$ to be indeed positive,
since $q^{2}+3\gamma>0$ and $q>0$ imply that $q-2p<0$.

\begin{figure}[ptb]
\includegraphics[height=4.5cm]{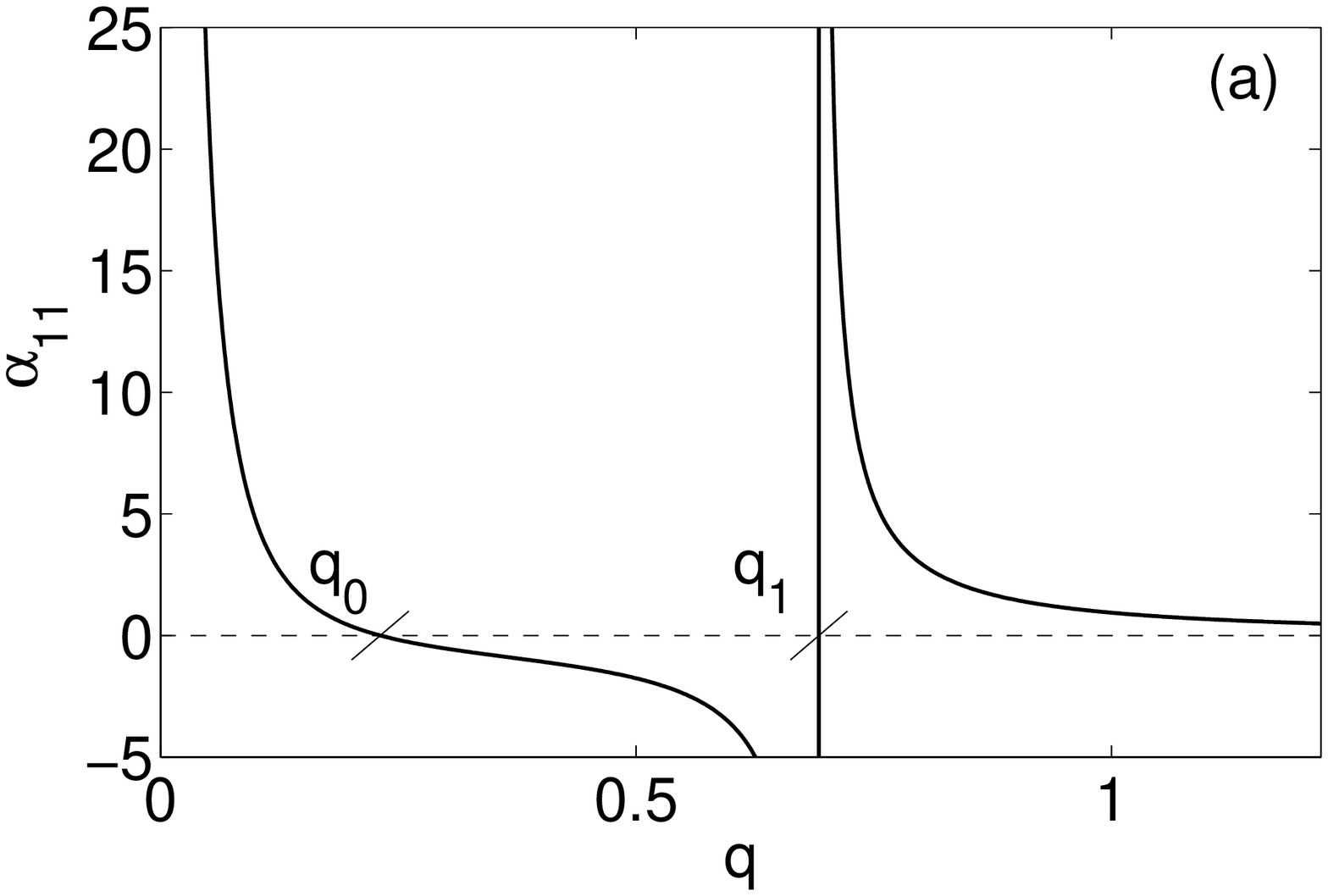}
\par
\includegraphics[height=4.5cm]{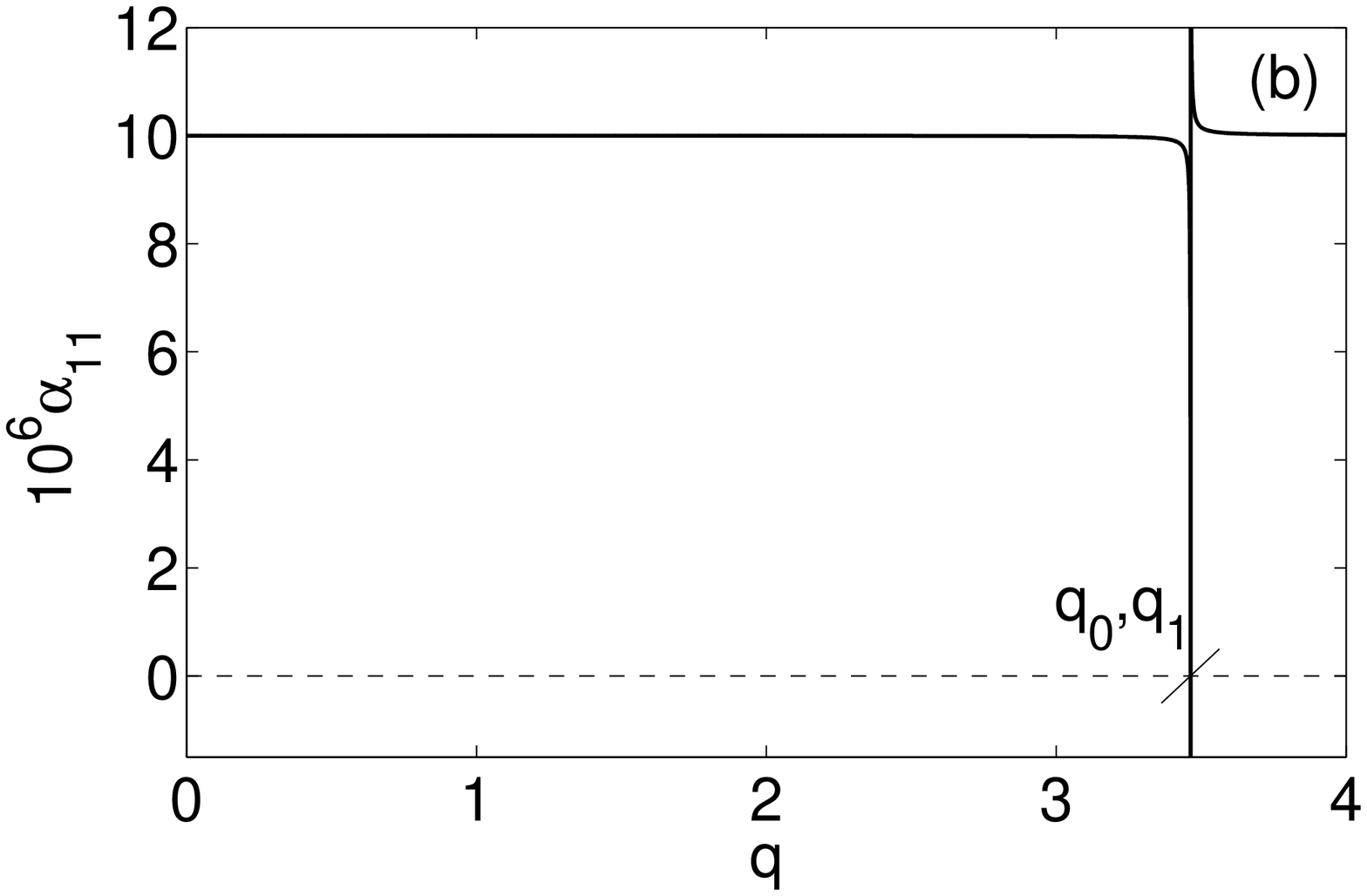}
\caption{The variation in $\protect\alpha_{11}$ as a function of
$q$, for (a) $\protect\gamma=0$ and (b) $\protect\gamma=10^{5}$.
Note that $ q_{0}<q_{1}$ and that
$q_{0},q_{1}\rightarrow2\protect\sqrt{3}$ for large $
\protect\gamma$.} \label{fig:a11_mondec_eva}
\end{figure}

Substituting the expression for $p$ into $p^{2}-3$, the requirement $
p^{2}-3<0$ can be written in the form
\begin{equation}
25q^{6}+(30\gamma-12)q^{4}+(9\gamma^{2}-72\gamma)q^{2}-108\gamma^{2}<0.
\label{polynomial-q1}
\end{equation}
It is clear that, as the left-hand side of the inequality is a
polynomial of even order, it can be satisfied if its largest real
root, $q_{1}$, is positive. By performing a numerical analysis of
this polynomial, we find that real positive solutions exist for
all $\gamma$, and that $q_{1}\rightarrow2 \sqrt {3}$ as
$\gamma\rightarrow\infty$. As in the case of the GVA solutions, we
restrict ourselves to the physically interesting subspace of
$\gamma>0$. In this case, there always exist one and only one real
positive solution $q_{1}$ to the above polynomial. Thus, provided
$0<q<q_{1}$ (i.e. $p^{2}-3<0$) and $q^{2}+3\gamma>0$, the
requirements that $p>0$, $Q>0$ and $P>0$ are met.

Turning to the remaining equation for $\alpha_{11}$, Eq. (\ref{eqn:qnew}),
with $\alpha_{11}\geq0$, we next find that we must have $4p^{3}-q(p^{2}
+3)\leq0$ (in conjunction with $p^{2}-3<0$) on the domain $0<q<q_{1}$. Here,
we have also taken into account the fact that the term $(q^{2}+\gamma)$ is
always positive, for $\gamma>0$. Substituting the expression for $p$ from
Eq. (\ref{eqn:p}) into $4p^{3}-q(p^{2}+3)\leq0$ we find that this inequality
is equivalent to
\begin{align}
& \left. 241q^{8}+(345\gamma-12)q^{6}+\gamma(171\gamma-108)q^{4}\right.
\notag \\
& \left. +\gamma^{2}(27\gamma-324)q^{2}-324\gamma^{3}\leq0\right. ,
\label{polynomial-q0}
\end{align}
and can be satisfied if the largest real root $q_{0}$ to the
polynomial in the left-hand side is positive. We again employ
numerics to find that for $\gamma>0$ there exist only one real and
positive root $q_{0}$, i.e. the above inequality is satisfied for
$q$ on $0<q\leq q_{0}$. In the limit of large $\gamma$,
$q_{0}\rightarrow2\sqrt{3}$.

Finally, as for the GVA solutions, we can show numerically (see Fig. \ref
{fig:a11_mondec_eva}) that $\alpha_{11}$ is a monotonically decreasing
function of $q$ on $0<q<q_{0}$, and that $q_{0}<q_{1}$. It approaches its
maximum value $1/\gamma$ as $q\rightarrow0$, and $\alpha_{11}=0$ at
$q=q_{0}$. This implies that there exists one-to-one correspondence between the
intervals $0<q\leq q_{0}$ and $0\leq\alpha_{11}<1/\gamma$ ($\gamma>0$),
so that the later can equally be regarded as the soliton existence
domain for the EVA solutions on the parameter space
($\alpha_{11}$,$\gamma$).


\begin{thebibliography}{99}
\bibitem{constrained-variables} The total number of particles $\mathcal{N}
^{\prime }=\mathcal{N}^{\prime }(P,Q,p,q)$ is a single valued function of
the EVA variables, and each of these variables can be treated as a single
valued function of $\mathcal{N}^{\prime }$ and the three remaining
variables. Because of this, the specific choice of the variables used
here for the
constrained minimization [relying on the elimination of $P$ in favor of $
\mathcal{N}^{\prime }$, via $P^{2}=p^{3}(\mathcal{N}^{\prime }/\pi
-2Q^{2}/q^{3})$] is just a matter of convinience and appears to be the
simplest.
\bibitem{Comment-spherical-symmetry} Calculations with enforced spherical
symmetry were used only after ensuring that the full $3D$ lattice
simulations give identical results, which was verified for a smaller set of
test points in the ($\alpha_{11}$-$\gamma$) parameter space.
\bibitem{Comment-time-scales} The relevant time scales for dynamical
stability of the soliton solutions, for each pair of the parameters $
\alpha_{11}$ and $\gamma$, are determined by comparing the results with
those observed for the case when the nonlinear coupling terms are removed
from the equations of motion.
\bibitem{Comment-trajectories} For example, defining vectors
$\vec{u}=(u,\sqrt{2}v)$, we interpolate between stable and
unstable solutions $\vec{u}_S$ and $\vec{u}_U$ by first setting
$\vec{u}^{\prime}_{\nu}=\vec{u}_S+\nu(\vec{u}_U-\vec{u}_S)$ where
$\nu$ runs from $0$ to $1$ and then normalizing so that
$\vec{u}_{\nu}=\sqrt{\mathcal{N}^{\prime }}\vec{u}^{\prime}_{\nu}/|\vec{u}^{\prime}_{\nu}|$.
The variable $s$ is an arc length defined as
$s(\vec{u}_{\nu})=\sqrt{\mathcal{N}^{\prime }}\cos^{-1}(|\vec{u}_{\nu}\cdot\vec{u}_S|/\mathcal{N}^{\prime })$.
Here, inner products are defined as
$\vec{u}_1\cdot\vec{u}_2=4\pi\int dr r^2(u_1^*u_2+2v_1^*v_2)$ so
that $\vec{u}\cdot\vec{u}=\mathcal{N}^{\prime}$.
\end{thebibliography}
\end{document}